\documentclass[a4paper,12pt]{article}
\usepackage{caption}
\PassOptionsToPackage{table}{xcolor}
\usepackage{jheppub} % for details on the use of the package, please
\usepackage[T1]{fontenc} % if needed

\usepackage[all]{xy}
\usepackage{orcidlink}
\usepackage[percent]{overpic}
\usepackage{slashed}
\usepackage{wrapfig}
\usepackage{tabu}
\usepackage{diagbox}
\usepackage{mathrsfs,amsmath,amssymb,amsthm,amsfonts,tikz,graphicx,accents,hyperref, color}
\usepackage{dsfont,epiolmec, latexsym, stmaryrd, comment}
\usepackage{slashed,ccaption}
\usepackage{mathrsfs, calligra}
\usepackage{leftidx}
\usepackage{import}
\usepackage{multirow}
\usepackage{amsfonts}
\usepackage{pifont}
\usepackage{tabularx}
\usepackage[utf8]{inputenc}
\usetikzlibrary{intersections,calc}
\usepackage{tikz-3dplot}
\usepackage{ifthen}
\usetikzlibrary{arrows}
\usepackage{amsmath}
\usepackage{xcolor}
\usetikzlibrary{backgrounds}

\usepackage{caption}
\usepackage{array}
\usepackage{tikz-3dplot}

\usepackage[percent]{overpic}
\usepackage{wrapfig}
\usepackage{bbm}
\usepackage{tabu}
\usepackage{slashed}
\usepackage{fancyhdr} % package for changing Headings style
\usepackage{amsmath}
\usepackage{amsfonts}
\usepackage{amssymb}
\usepackage{diagbox}

\usepackage{multirow}
\usepackage{pifont}% http://ctan.org/pkg/pifont

\hypersetup{ linktoc=all,
    colorlinks, linkcolor={blue}, %{vividviolet},
    citecolor={red}, urlcolor={darkpink}
}

\graphicspath{{Images/}}

\definecolor{light gray}{RGB}{220,220,220}
\definecolor{dark purple}{RGB}{108,0,217}
\definecolor{pink}{RGB}{190,20,100}
\definecolor{orang}{RGB}{193,63,0}
\definecolor{green}{RGB}{11,98,17}
\definecolor{darkpink}{RGB}{153,0,76}
\definecolor{bluegreen}{RGB}{0,102,102}
\definecolor{greenlagan}{RGB}{0,102,0}
\definecolor{redgreen}{RGB}{102,102,0}
\definecolor{Redgreen}{RGB}{153,76,0}
\definecolor{vividviolet}{rgb}{0.62, 0.0, 1.0}
\definecolor{amaranth}{rgb}{0.9, 0.17, 0.31}
\definecolor{palatinateblue}{rgb}{0.15, 0.23, 0.89}
\definecolor{brightpink}{rgb}{1.0, 0.0, 0.5}
\definecolor{cornflowerblue}{rgb}{0.39, 0.58, 0.93}
\definecolor{deepcarminepink}{rgb}{0.94, 0.19, 0.22}
\definecolor{radicalred}{rgb}{1.0, 0.21, 0.37}
\usepackage{graphicx}
\usepackage{tikz}
\usetikzlibrary{arrows,chains,shapes,matrix,positioning,scopes}
\usepackage{leftidx}
\usepackage{float}
\usetikzlibrary{decorations.markings}
\tikzstyle arrowstyle=[scale=1]
\tikzstyle directed=[postaction={decorate,decoration={markings,
    mark=at position .65 with {\arrow[arrowstyle]{stealth}}}}]
\tikzstyle reverse directed=[postaction={decorate,decoration={markings,
    mark=at position .65 with {\arrowreversed[arrowstyle]{stealth};}}}]
\usepackage{import}
\usepackage{accents}
\usepackage{mathrsfs,amsmath,amssymb,slashed}
\usepackage{multirow,multicol}
\usepackage{enumitem}
\usepackage[percent]{overpic}
\usepackage{slashed}

\usepackage{wrapfig}
\usepackage{tabu}
\usepackage{diagbox}
\usepackage{comment}
\usepackage{tikz}
\usepackage{mathrsfs,amsmath,amssymb,amsthm,amsfonts,graphicx,accents,hyperref,color}
\usepackage{leftidx}
\usepackage{import}
\usetikzlibrary{decorations.pathmorphing}
\DeclareFontFamily{OT1}{rsfs}{}

\DeclareFontShape{OT1}{rsfs}{m}{n}{ <-7> rsfs5 <7-10> rsfs7 <10->rsfs10}{} 

\DeclareMathAlphabet{\mycal}{OT1}{rsfs}{m}{n}

\newcommand{\be}{\begin{equation}}
\newcommand{\ee}{\end{equation}}

    \title{Carrollian Quantum Mechanics:\\ Time-like, Space-like and Hybrid Sectors}
    
\author{\, Mehdi~Ahmadi-Jahmani \orcidlink{0009-0008-0637-8764}}
\affiliation{Department of Physics, Faculty of Science, Ferdowsi University of Mashhad, Mashhad, Iran}
\emailAdd{mehdiahmadijahmani@gmail.com}

\abstract{
We develop a comprehensive theoretical framework for Carrollian quantum mechanics by performing systematic ultra-relativistic contractions of the Klein–Gordon equation in the limit \(c \to 0\). This limiting process uncovers three distinct sectors—time-like, space-like, and hybrid—each governed by a Carroll-invariant wave equation and accompanied by a consistent probabilistic interpretation. The time-like sector exhibits a novel temporal tunneling phenomenon, characterized by the relation \(|\mathcal{T}|^2 = 1 + |\mathcal{R}|^2\), which reflects the indefinite character of the Klein–Gordon norm. In the space-like sector, the presence of spatial propagation necessitates a tachyonic dispersion relation, which forces the density to vanish for energy eigenstates, yielding zero-norm (null) states that gain physical relevance within Carrollian physics due to the underlying null geometric structure. The hybrid sector combines features of both sectors and admits two distinct formulations, which may be either tachyonic or non-tachyonic, and can be interpreted as an inhomogeneous Klein–Gordon equation. For all three sectors, we derive the corresponding continuity equations, probability densities, and currents, and investigate canonical quantum systems—including the particle in a box and tunneling phenomena—within the Carrollian regime.
}

\begin{document}
\maketitle
%%%%%%%%%%%%%%%%%%%%%%%%%%%%%%%%%%%%%%%%%%%%%%%%%%%
\section{Introduction}

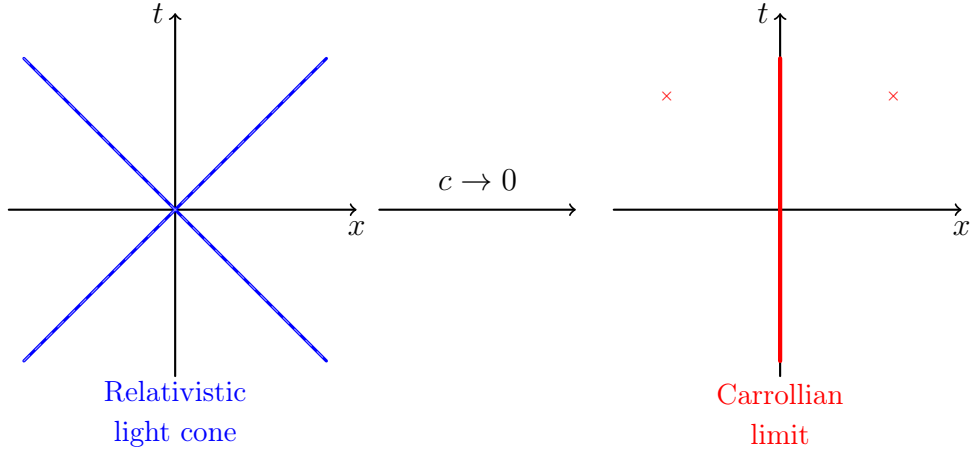
\begin{figure}[t]
	\centering
	\begin{tikzpicture}[scale=1, line cap=round, line join=round]
		
		% ================= Relativistic light cone =================
		\begin{scope}[shift={(-4,0)}]
			
			% Axes
			\draw[->, thick] (0,-2.2) -- (0,2.6) node[left] {$t$};
			\draw[->, thick] (-2.2,0) -- (2.4,0) node[below] {$x$};
			
			% Light cone
			\draw[very thick, blue] (-2,-2) -- (0,0) -- (2,-2);
			\draw[very thick, blue] (-2,2) -- (0,0) -- (2,2);
			
			% Lightlike guides (optional)
			\draw[dashed, blue!40] (-2,-2) -- (2,2);
			\draw[dashed, blue!40] (-2,2) -- (2,-2);
			
			% Label
			\node[blue, font=\small, align=center] at (0,-2.7) {Relativistic\\ light cone};
		\end{scope}
		
		% ================= Limit arrow =================
		\draw[->, thick] (-1.3,0) -- (1.3,0) node[midway, above=3pt] {$c \rightarrow 0$};
		
		% ================= Carrollian limit =================
		\begin{scope}[shift={(4,0)}]
			
			% Axes
			\draw[->, thick] (0,-2.2) -- (0,2.6) node[left] {$t$};
			\draw[->, thick] (-2.2,0) -- (2.4,0) node[below] {$x$};
			
			% Collapsed cone (time axis only)
			\draw[ultra thick, red] (0,-2) -- (0,2);
			
			% Label
			\node[red, font=\small, align=center] at (0,-2.7) {Carrollian\\ limit};
			
			% Optional: add small "x" symbol to indicate spatial degeneracy
			\node[red, font=\tiny] at (1.5,1.5) {$\times$};
			\node[red, font=\tiny] at (-1.5,1.5) {$\times$};
			
		\end{scope}
		
	\end{tikzpicture}
	\caption{Geometric illustration of the Carrollian limit. As \(c \to 0\), the relativistic light cone (blue) collapses onto the time axis (red). Spatial propagation is eliminated: time‑like Carroll particles become immobile, while space‑like and hybrid sectors retain constrained spatial dynamics. The degenerate causal structure is the hallmark of Carrollian geometry.}
	\label{fig:carroll_limit}
\end{figure}

Carrollian physics has emerged as a prominent and well-established framework for exploring phenomena in high-energy theoretical physics. This framework is obtained by taking the ultra-relativistic limit \(c \to 0\) from relativistic physics (see Fig.~\ref{fig:carroll_limit})~\cite{levy1965nouvelle,sen1966analogue}. The growing interest in Carrollian physics is largely due to its applicability in diverse areas such as flat space holography, null surface dynamics, fracton phases, Carrollian gravity, and tensionless string theories~\cite{Bergshoeff:2022qkx,Ciambelli:2019lap,Donnay:2019jiz,Ecker:2023uwm,Herfray:2021qmp,Adami:2023wbe,Bagchi:2010zz,Ciambelli:2018wre,Nguyen:2023carrollian,Bagchi:2023cen,Ahmadi-Jahmani:2025iqc,Figueroa-OFarrill:2023vbj,Figueroa-OFarrill:2023qty,Bidussi:2021nmp,cardona2016dynamics,Bagchi:2019cay,Bagchi:2019tensionless,Afshar:2025imp,Hartong:2015xda,Bergshoeff:2017btm}. 

Just as the Poincaré group leads to the Klein-Gordon equation and the Galilei group to the Schrödinger equation, it is natural to ask what quantum equation is dictated by the Carroll group~\cite{Marsot:2022imf}. Despite this basic question, a systematic, self-consistent quantum mechanical formulation of the Carroll limit has been missing. Prior works derived isolated equations (e.g., the time-like Carroll equation~\cite{Marsot:2022imf}) or remained at the classical level~\cite{Figueroa-OFarrill:2023vbj,Ahmadi-Jahmani:2025iqc}. However, no unified treatment exists for the three inequivalent Carrollian sectors—time-like, space-like, and hybrid—that emerge from a controlled contraction of the Klein-Gordon equation. Moreover, the probabilistic interpretation (continuity equation, probability density, and current) has not been established for each sector, and novel phenomena such as \emph{temporal tunneling}—the mixing of positive and negative frequencies across a time-dependent barrier—have never been explored.

In this work, starting from the Klein-Gordon equation, we perform systematic Carrollian contractions that yield three inequivalent quantum theories, each invariant under Carrollian transformations. The first, well-defined contraction leads to time-like Carroll quantum mechanics, in which only temporal derivatives survive. Particles in this sector have no spatial dynamics—the probability current vanishes—yet they exhibit a purely \emph{temporal tunneling} effect. We solve the temporal infinite well, compute reflection and transmission amplitudes for a rectangular temporal barrier, and obtain \(|\mathcal{T}|^2 = 1 + |\mathcal{R}|^2\), a direct signature of the indefinite Klein-Gordon norm.

A second contraction preserves spatial derivatives, giving rise to the space-like Carrollian sector. This sector is not invariant under Carrollian boosts in its naive form; to restore invariance, an additional compensating term must be introduced. This term can take two forms: a first-order temporal derivative coupled to a field $\chi$ (Approach 1), which yields tachyonic null states, or an additive term $\tilde{\chi}$ without derivatives (Approach 2), which describes non-tachyonic configurations.

The third resulting sector is the hybrid Carrollian sector, where both temporal and spatial derivatives appear. As in the space-like case, an additional term is required to restore Carroll boost invariance. This term can take two forms: a first-order temporal derivative (Approach 1), producing a damped or amplified wave equation with complex energy eigenvalues—tachyonic for complex wavefunctions and non-tachyonic for real wavefunctions—or an additive term devoid of derivatives (Approach 2), reducing to an inhomogeneous Klein-Gordon equation with real energy spectrum. We analyze the particle in a box spectrum—identifying underdamped, critically damped, and overdamped regimes—and solve both spatial and temporal barrier scattering.

For each sector, we provide a consistent probabilistic interpretation (continuity equation, positive-definite density where applicable, and probability current). We also obtain exact solutions for the particle in a box and rectangular barrier problems, including explicit scattering amplitudes. Finally, we establish a correspondence between time-like Carroll particles and fractonic monopoles, linking immobile excitations in Carrollian quantum mechanics to fracton phases \cite{Figueroa-OFarrill:2023qty}. Our results therefore supply a unified and predictive quantum framework relevant to flat space holography, null surface quantization, non-Lorentzian field theory, and fracton physics.

The paper is organized as follows. Section~\ref{Relativistic Spin-0 Particles: Klein-Gordon Equation} reviews the Klein-Gordon equation. Section~\ref{Carroll sym} defines the three Carrollian limits. Sections~\ref{sec:timelike}, \ref{sec:spacelike}, and \ref{sec:hybrid} analyze the time-like, space-like, and hybrid sectors, respectively. Section~\ref{conclusion} concludes the paper. The appendix presents the classical dynamics of hybrid Carrollian particles.

%%%%%%%%%%%%%%%%%%%%%%%%%%%%%%%%%%%%%%%%%%%%%%%%%%%%%%%%%%%%%%%%%%%%%%%%%%%%%%%%%%%%%%%%%%%%%%%%%%%%%%%%%%%%%%%%%%%%%%%%%%%%%%%%%%%%%%%%%%%%%%%%%%%%%%%%%%%%%%%%%%%%%%%%%%%%%%%%
\section{Relativistic Spin-0 Particles: Klein-Gordon Equation}\label{Relativistic Spin-0 Particles: Klein-Gordon Equation}

Investigating high-energy phenomena demands relativistic wave equations that remain invariant under Lorentz transformations \cite{BjorkenDrell,GreinerRQM,SakuraiAQM}. Here we consider spin-$0$ quantum particles. The Klein-Gordon equation provides the basic framework for massive \(m_0\) spin-$0$ particles. This equation encapsulates the relativistic energy–momentum relation and is formulated as:
\begin{align}\label{Klein-Gordon Equation}
	\left(\frac{1}{c^2}\partial_t^2 - \partial_{x}^2 + \frac{m_0^2 c^2}{\hbar^2}\right)\psi(\vec{x}, t) = 0\,.
\end{align}
The fundamental solutions are plane monochromatic waves:
\begin{align}
	\psi(\vec{x}, t) \propto \exp\!\left(i \vec{k} \cdot \vec{x} - i\omega t\right).
\end{align}
These correspond to the eigenvalues
\begin{align}
	\vec{p} = \hbar \vec{k}, \qquad 
	E_{\pm} = \hbar \omega_{\pm} = \pm \sqrt{(\hbar c \vec{k})^2 + (m_0 c^2)^2}\,.
\end{align}
Thus, the Klein-Gordon equation admits both positive and negative energy solutions. The negative energy solutions are physically interpreted as antiparticles, and their experimental observation motivates the extension from non-relativistic to relativistic quantum mechanics.

A complex wavefunction \(\psi(\vec{x},t)\) endows the theory with a global \(U(1)\) phase symmetry, which in the context of charged spin-$0$ particles (e.g., pions) is interpreted as electric charge. In electrodynamics, charge conservation is expressed by the continuity equation
\begin{align}
	\partial_t \rho_e + \vec{\partial_{x}}\cdot \vec{J}_e = 0,
\end{align}
where \(\rho_e\) is the charge density and \(\vec{J}_e\) the current density. By analogy, we seek a continuity equation for probability. Starting from the Klein-Gordon equation and its complex conjugate,
\begin{align}
	\left(\frac{1}{c^2}\partial_t^2 - \partial_{x}^2 + \frac{m_0^2 c^2}{\hbar^2}\right)\psi &= 0, \\
	\left(\frac{1}{c^2}\partial_t^2 - \partial_{x}^2 + \frac{m_0^2 c^2}{\hbar^2}\right)\psi^* &= 0,
\end{align}
multiply the first by \(\psi^*\) and the second by \(\psi\), subtract, and obtain
\begin{align}
	\psi^*\big(\frac{1}{c^2}\partial_t^2 - \partial_{x}^2 + \frac{m_0^2 c^2}{\hbar^2}\big)\psi - \psi\big(\frac{1}{c^2}\partial_t^2 - \partial_{x}^2 + \frac{m_0^2 c^2}{\hbar^2}\big)\psi^* = 0.
\end{align}
This can be rearranged into the continuity equation
\begin{align}
	\partial_t \rho + \vec{\partial_{x}}\cdot \vec{J} = 0,
\end{align}
with the probability density and current given by
\begin{align}
	\rho(\vec{x},t) &= \frac{i\hbar}{2m_0 c^2}\big(\psi^*\partial_t\psi - \psi\partial_t\psi^*\big), \label{KG probability1}\\
	\vec{J}(\vec{x},t) &= \frac{i\hbar}{2m_0}\big(\psi^*\vec{\partial_{x}}\psi - \psi\vec{\partial_{x}}\psi^*\big).
\end{align}
For a plane wave, \(\rho\) becomes
\begin{align}
	\rho(\vec{x},t) = \pm \frac{E}{2m_0 c^2}\, \psi^*\psi, \label{KGprobability2}
\end{align}
where the sign depends on whether the solution has positive or negative energy. The appearance of a negative probability density for negative energy states is the well-known problem of the Klein-Gordon equation \cite{BjorkenDrell,GreinerRQM}. This forces the interpretation of negative energy solutions as antiparticles and ultimately requires a transition to quantum field theory, where particle number is not conserved.

Nevertheless, spin-$0$ particles such as pions are successfully described by the Klein-Gordon equation in relativistic quantum mechanics, despite its probabilistic shortcomings. In this work, we will apply a limiting procedure to the Klein-Gordon equation – the Carrollian (ultra-relativistic) contraction \(c\to0\)  – to obtain consistent quantum mechanical equations that avoid the indefinite probability issue in their respective physical sectors. These Carrollian quantum equations will be the subject of the following sections.
%%%%%%%%%%%%%%%%%%%%%%%%%%%%%%%%%%%%%%%%%%%%%%%%%%%%
\section{Carrollian Domain: From Relativistic to Carrollian} \label{Carroll sym}

Carroll symmetry is a non-Lorentzian spacetime symmetry that emerges from the Poincaré group via a specific Inönü-Wigner contraction: the limit where the speed of light \(c\) goes to zero~\cite{levy1965nouvelle}. Unlike the more familiar Galilean contraction (\(c\to\infty\)), which yields non-relativistic physics, the Carroll contraction (\(c\to0\)) produces an ultra-relativistic regime in which the light cone collapses onto the time axis, resulting in a degenerate causal structure: spatial propagation is frozen while temporal evolution remains (see Fig.~\ref{fig:carroll_limit}). The name \textbf{Carroll} was chosen by Lévy-Leblond and Sen-Gupta \cite{levy1965nouvelle,sen1966analogue} as a playful reference to Lewis Carroll's Through the Looking Glass, because the Red Queen's remark, "it takes all the running you can do to keep in the same place," aptly captures the immobility of Carroll particles \cite{levy1965nouvelle,Bergshoeff:2014jla,Ahmadi-Jahmani:2025iqc}.

Despite the seemingly unphysical nature of the \(c\to0\) limit, Carrollian symmetries have found numerous applications in modern theoretical physics: they govern the dynamics on null surfaces (black hole horizons, null infinity) \cite{Ciambelli:2019lap,Donnay:2019jiz,Ecker:2023uwm,Herfray:2021qmp,Adami:2023wbe}, appear in flat space holography as the symmetry of the boundary Carrollian conformal field theory \cite{Bagchi:2010zz,Ciambelli:2018wre,Nguyen:2023carrollian,Bagchi:2023cen,Afshar:2024llh}, describe fractonic excitations in condensed matter \cite{Ahmadi-Jahmani:2025iqc,Figueroa-OFarrill:2023vbj,Figueroa-OFarrill:2023qty,Bidussi:2021nmp}, and characterize the world-sheet of tensionless (null) strings \cite{cardona2016dynamics,Bagchi:2019cay,Bagchi:2019tensionless}. A systematic quantum mechanical formulation of Carrollian physics is, therefore, not only a mathematical curiosity but a necessary step toward understanding these diverse areas.

\subsection{Carroll boosts and the degenerate metric}

The Carroll group can be obtained by contracting the Poincaré group. In physical terms, we consider the limit \(c\to0\) via a dimensionless parameter \(\epsilon\):
\[
c \to \epsilon c,\qquad \epsilon \to 0.
\]
After taking the limit, we may set \(c=1\) by rescaling units, but the degenerate structure of the metric remains. The finite Carroll boosts and rotations act on coordinates as~\cite{Bergshoeff:2022eog,deBoer:2021jej,Ahmadi-Jahmani:2025iqc}
\begin{align}
	\vec{x}' &= R\,\vec{x}, \\
	t' &= t - \vec{\beta}\cdot\vec{x},
\end{align}
where \(\vec{\beta}\) is the Carroll boost parameter. Under this transformation, spatial derivatives transform as \(\vec{\partial}' = \vec{\partial} + \vec{\beta}\,\partial_t\), while the temporal derivative remains invariant, \(\partial_t' = \partial_t\). Higher derivatives transform accordingly~\cite{Afshar:2025imp}:
\begin{align}
	\partial_i'\partial_i' &= \partial_i\partial_i + 2\beta_i\partial_i\partial_t + \beta^2\partial_t^2, \\
	\partial_i'\partial_t' &= \partial_i\partial_t + \beta_i\partial_t^2.
\end{align}
Thus, the Carrollian structure is characterized by a degenerate metric: the time direction becomes null (light-like), while the spatial directions remain absolute (i.e., they carry a non-degenerate Euclidean metric). This is the ultra-relativistic counterpart of Galilean geometry, where time is absolute and space is degenerate. The causal structure therefore allows only propagation along the time direction; spatial motion is frozen, which is the origin of the immobility of time-like Carroll particles.
%---------------------------------------------------------------------
%-----------------------------------------------------------------
\subsection{Three Carrollian limits from the Klein-Gordon equation}\label{Carrollian limits}

To construct Carrollian quantum mechanics, we apply the above contraction procedure directly to the Klein-Gordon equation \eqref{Klein-Gordon Equation}. Depending on \emph{which} variables are rescaled and how the wavefunction is scaled, we obtain three physically distinct quantum theories. Each corresponds to a different way of “sitting” on the degenerate light cone. The two limits (time-like and space-like) were first identified at the classical level in~\cite{Ahmadi-Jahmani:2025iqc}\footnote{The classical hybrid limit is not mentioned in any paper; in Appendix~\ref{app:classical_hybrid} we obtain this classical limit in a consistent manner.}; here we adapt them to quantum mechanics.

\paragraph{Time-like Carroll limit (electric sector).}  
This limit retains the temporal derivative but discards spatial derivatives. It is achieved by the scaling
\begin{align}\label{time-like Carroll limit}
	m \to \frac{m}{\epsilon^2}, \qquad \psi \to \epsilon^2 \psi, \qquad \epsilon \to 0.
\end{align}
Physically, this describes particles that are \emph{completely immobile}---they cannot propagate in space but can evolve in time. As we will see in Sec.~\ref{sec:timelike}, such particles admit a consistent probabilistic interpretation (positive probability density) when restricted to positive frequencies. They are the quantum counterparts of the time-like Carroll particles discussed in~\cite{Ahmadi-Jahmani:2025iqc} and are dual to \emph{isolated fractons (monopoles)} \cite{Figueroa-OFarrill:2023qty}.

\paragraph{Space-like Carroll limit (magnetic sector).}  
This limit preserves spatial derivatives but removes the second-order time derivative. It requires a compensating field to maintain Carroll invariance, and the scaling is
\begin{align}\label{space-like Carroll limit}
	t \to \frac{t}{\epsilon},\quad \vec{x} \to \epsilon \vec{x},\quad m \to \frac{m}{\epsilon},\quad \psi \to \epsilon^{2}\psi,\quad \epsilon\to 0.
\end{align}
The resulting equation is first order in time, analogous to the Schrödinger equation, but with the temporal derivative coupled to a compensating field \(\chi\). As we will show in Sec.~\ref{sec:spacelike}, this sector admits two distinct formulations: a coupled formulation yielding tachyonic null states, and an additive formulation describing non-tachyonic static configurations. In the classical limit, these space-like Carroll particles can propagate spatially \cite{Ahmadi-Jahmani:2025iqc}.

\paragraph{Hybrid Carroll limit.}  
This limit retains \emph{both} temporal and spatial derivatives, albeit with a modified structure that includes a first-order time derivative term. The scaling is
\begin{align}\label{hybrid Carroll limit}
	\vec{x} \to \epsilon \vec{x},\quad m \to \frac{m}{\epsilon},\quad \psi \to \epsilon^{2}\psi,\quad \epsilon\to 0.
\end{align}
Like the previous two, the hybrid limit admits a classical analogue, which we present in Appendix~\ref{app:classical_hybrid}. As we will show in Sec.~\ref{sec:hybrid}, the corresponding quantum wave equation admits two formulations: a coupled formulation yielding a damped or amplified wave equation with tachyonic behavior for complex wavefunctions, and an additive formulation reducing to a non-tachyonic inhomogeneous Klein-Gordon equation. This sector interpolates between time-like and space-like behaviors and may be relevant for understanding Carroll swiftons~\cite{Ecker:2024czx}, though a detailed investigation is left for future work.

%%%%%%%%%%%%%%%%%%%%%%%%%%%%%%%%%%%%%%%%%%%%%%%%%%%%%%%%%%%%%%%%%%%%%%%%%%%%%%%%%%%%%%%%%%%%%%%%%%%%%%%%%%%%%%%%%%%%%%%%%%%%%%%%%%%%%%%%%%%%%%%%%%%%%%%%%%%%%%%%%%%%%%%%%%%%%%%%%%%%
\section{Time-like Carroll QM}\label{sec:timelike}

Classical time-like Carroll particles are completely immobile: they cannot move in space, yet they possess electric charge and respond non-trivially to electromagnetic fields \cite{Ahmadi-Jahmani:2025iqc,Bergshoeff:2022qkx,Bergshoeff:2014jla,Bergshoeff:2022eog}. Their quantum description is obtained by applying the time-like contraction \eqref{time-like Carroll limit} to the Klein-Gordon equation \eqref{Klein-Gordon Equation}. This yields the time-like Carroll wave equation (originally introduced by Marsot \cite{Marsot:2021tvq})
\begin{align}\label{time-like Carroll QM}
	(\hbar^{2}\partial_{t}^{2}+m_{0}^{2})\psi(\vec{x},t)=0,
\end{align}
which contains only second order time derivatives; spatial derivatives are completely absent. In the classical theory, such particles are dual to isolated fractons \cite{Ahmadi-Jahmani:2025iqc,Figueroa-OFarrill:2023vbj}; the quantum equation \eqref{time-like Carroll QM} therefore provides a direct link to fractonic excitations.

\subsection{Carroll invariance and complex wavefunction}
Equation \eqref{time-like Carroll QM} is invariant under Carroll transformations. For a real field, the wavefunction transforms as a scalar: $\psi(\vec{x},t)=\psi'(\vec{x}',t')$. To describe charged particles we promote $\psi$ to a complex field, with its complex conjugate satisfying the same equation:
\begin{align}\label{time-like complex conjugate}
	(\hbar^{2}\partial_{t}^{2}+m_{0}^{2})\psi^{*}(\vec{x},t)=0.
\end{align}
The global $U(1)$ symmetry of the complex theory is consistent with the interpretation of time-like Carroll particles as \emph{electric Carroll particles} \cite{Ahmadi-Jahmani:2025iqc}.

\subsection{Probability conservation}
Following the analogy with electrodynamics, we construct a continuity equation. From \eqref{time-like Carroll QM} and its conjugate \eqref{time-like complex conjugate}, 
we obtain
\begin{align}
    \psi^{*}(\hbar^{2}\partial_{t}^{2}+m_{0}^{2})\psi 
    - \psi(\hbar^{2}\partial_{t}^{2}+m_{0}^{2})\psi^{*}=0,
\end{align}
which can be rewritten as
\begin{align}
    \partial_{t}\rho(\vec{x},t) + \partial_{x}\cdot\vec{J}(\vec{x},t)=0,
\end{align}
with
\begin{align}
    \rho(\vec{x},t) &= i\hbar\big(\psi^{*}\partial_{t}\psi - \psi\partial_{t}\psi^{*}\big), \\
    \vec{J}(\vec{x},t) &= 0.
\end{align}
The vanishing current is a direct consequence of the absence of spatial derivatives: probability cannot flow between spatial points. Each spatial point evolves independently, which is precisely the quantum signature of fractonic immobility.

\paragraph{Probability density and physical Hilbert space in the time-like sector.}
The time-like Carroll equation inherits a Klein-Gordon-type conserved density, which is not positive-definite for arbitrary solutions. Because the equation is second order in time, it admits both positive and negative frequency modes, and the density $\rho$ changes sign between these branches. Negative-frequency solutions correspond to negative rest energy and yield $\rho<0$, which is incompatible with a probabilistic interpretation. Moreover, in our single-particle Carrollian framework, the Carroll group acts only on spacetime and does not supply a charge-conjugation symmetry that could reinterpret the negative-frequency branch as physical antiparticles. Such states are therefore excluded from the physical Hilbert space.

We define the physical Hilbert space by restricting to the positive-frequency sector, which is closed under Carrollian time evolution and yields a positive-definite probability density,
\begin{align}
    \rho = 2E_{0}|\psi|^{2} > 0,
\end{align}
providing a consistent Born-rule interpretation.

This restriction is standard: it is the direct Carrollian analogue of the positive-energy quantization procedure in relativistic quantum mechanics, where the positive-energy branch defines the physical one-particle Hilbert space.

%--------------------------------------------------------------------------------------------------------------------------------------------------------------------------------------

%-------------------------------------------------

%-------------------------------------------------

\subsection{Time-like Carrollian particle in a temporal box}\label{subsec:timelike_box}

Because time-like Carroll particles have no spatial dynamics, it is natural to consider confining them in \emph{time} rather than in space. A temporal box is defined by a time dependent potential \(U(t)\) that is constant (or zero) inside an interval \(t_1<t<t_2\) and infinite outside, forcing the wavefunction to vanish at the two boundaries:
\begin{align*}
	U(t) = 
	\begin{cases}
		\infty, & t < t_1,\\
		V\ \ (\text{constant}), & t_1 \le t \le t_2,\\
		\infty, & t > t_2,
	\end{cases}
\end{align*}
where \(V\) is a real constant. The time-like Carroll equation inside the box is
\begin{align}
	\hbar^2\partial_t^2\psi(x,t) + \bigl(m_0^2 + V^2\bigr)\psi(x,t)=0,
\end{align}
which is independent of the spatial coordinate \(x\) (no spatial derivatives). Hence \(\psi(x,t)=\phi(x)\xi(t)\) with \(\phi(x)\) arbitrary (subject to spatial boundary conditions) and \(\xi(t)\) satisfying
\begin{align}
	\ddot{\xi} + \omega^2 \xi = 0,\qquad \omega = \frac{\sqrt{m_0^2+V^2}}{\hbar}.
\end{align}
The infinite walls impose \(\xi(t_1)=\xi(t_2)=0\). Shifting the origin to \(t_1\) (\(s=t-t_1\)), we have
\begin{align*}
	\xi(s)=A\sin(\omega s)+B\cos(\omega s).
\end{align*}
The condition \(\xi(0)=0\) gives \(B=0\); then \(\xi(T)=A\sin(\omega T)=0\) with \(T=t_2-t_1\). Thus \(\omega T = n\pi\) for \(n=1,2,\dots\). This yields the master quantization condition
\begin{align}
	\sqrt{m_0^2 + V^2}\;T = n\pi\hbar.
\end{align}

\paragraph{Case 1: Zero interior potential (\(V=0\)).}
The master condition becomes \(m_0 T = n\pi\hbar\), i.e.
\begin{align}
	m_0 = \frac{n\pi\hbar}{T}.
\end{align}
The rest mass of the Carroll particle is \textbf{quantized} in units of \(\pi\hbar/T\). For a given box duration \(T\), only particles with specific masses can be confined. Conversely, for a particle of fixed mass \(m_0\), the box duration must be \(T = n\pi\hbar/m_0\). This mathematically consistent prediction is unconventional because the mass depends on an external parameter, indicating that the zero potential temporal box is an idealized limiting case.

\paragraph{Case 2: Non-zero constant interior potential (\(V\neq0\)).}
Here the particle mass \(m_0\) is fixed, and the master condition quantities the potential depth:
\begin{align}
	V_n = \pm\sqrt{\left(\frac{n\pi\hbar}{T}\right)^2 - m_0^2}.
\end{align}
The effective oscillation frequency is quantized:
\begin{align}
	\omega_n = \frac{n\pi}{T},\qquad \hbar\omega_n = \frac{n\pi\hbar}{T}.
\end{align}
Thus the temporal box acts as a resonant cavity in time: only specific values of \(|V|\) (or equivalently specific effective energies) satisfy the boundary conditions. For a fixed mass \(m_0\) and box duration \(T\), there is a minimum \(n\) such that \(n\pi\hbar/T > m_0\) for real \(V\); below that threshold, \(V\) becomes imaginary, indicating that the particle cannot be confined.

\paragraph{Full wavefunctions and probability density.}
The normalized temporal modes for both cases are
\begin{align}
	\xi_n(t) = \sqrt{\frac{2}{T}}\sin\!\left(\frac{n\pi(t-t_1)}{T}\right),\qquad t_1\le t\le t_2,
\end{align}
and zero outside. The full wavefunction is \(\psi_n(x,t)=\phi(x)\xi_n(t)\). The probability density is
\begin{align}
	\rho_n(x,t) = 2m_0|\psi_n(x,t)|^2 = 2m_0|\phi(x)|^2\,\xi_n(t)^2,
\end{align}
which is positive and time dependent. The spatial distribution \(|\phi(x)|^2\) is frozen because the probability current vanishes; only the temporal envelope changes.

\paragraph{Physical interpretation.}
The temporal box illustrates how the degenerate causal structure of Carrollian physics leads to quantization in time rather than in space. When \(V=0\) the rest mass is quantized; when \(V\neq0\) the effective oscillation energy is quantized while the mass remains fixed. Since the wave equation depends only on \(V^2\), a potential barrier (\(V>0\)) and a potential well (\(V<0\)) of the same magnitude produce identical physical predictions  the \(\pm\) sign is irrelevant. This quadratic coupling is a distinctive feature of Carrollian quantum mechanics, contrasting with the linear coupling of potentials in the Schrödinger equation.

\paragraph{Uncertainty principle and immobility.}
The time-like Carroll equation contains no spatial derivatives, so the spatial profile $\phi(x)$ of the wavefunction $\psi(x,t)=\phi(x)e^{-iE_0 t/\hbar}$ is arbitrary (subject to normalization). Immobility in this sector means that the probability current vanishes and the spatial probability density $|\psi(x,t)|^2$ is time-independent. However, the wavefunction may still have a finite spatial spread ($\Delta x \neq 0$), since $\phi(x)$ can be localized or broad. The momentum operator $\hat{p}=-i\hbar\partial_x$ acts only on $\phi(x)$ and satisfies the canonical commutation relation $[x,p]=i\hbar$; therefore the Heisenberg uncertainty relation $\Delta x\,\Delta p \ge \hbar/2$ holds for any state. The Hamiltonian $H=m_0$ commutes with $\hat{p}$, so momentum eigen-states are stationary, but this does not restrict the possible momentum uncertainties, which are determined solely by $\phi(x)$. There is no contradiction: immobility refers to the absence of spatial probability flow, not a classical notion of rest.
%-------------------------------------------------------------------
%-------------------------------------------------------------------
\subsection{time-like tunneling}\label{subsec:timelike_tunneling}

The temporal box and barrier problems in the time-like Carroll sector admit a striking  formal analogy with standard non-relativistic scattering. The temporal boundaries act  as potential steps in the time domain: matching the wavefunction and its first time  derivative at the interfaces $t_1$ and $t_2$ yields reflection and transmission  amplitudes in complete analogy with spatial scattering off a rectangular potential. This correspondence allows us to interpret the mixing of positive and negative  frequencies as a temporal analogue of partial reflection and transmission. 

The key departure from ordinary quantum mechanics lies in the underlying norm: because  the time-like Carroll equation is second-order in time with an indefinite inner  product, the Bogoliubov relation
\begin{equation}
    |\mathcal{T}|^{2} = 1 + |\mathcal{R}|^{2}
    \label{eq:bogoliubov_temporal}
\end{equation}
replaces the usual unitarity condition $|\mathcal{T}|^{2}+|\mathcal{R}|^{2}=1$ of  Schr\"{o}dinger scattering. Thus, the scattering theory framework provides both a  robust computational tool and a valuable conceptual bridge for analyzing temporal  tunneling phenomena. 

Consider a time dependent potential \(U(t)\) (the spatial coordinate \(x\) plays no dynamical role because the equation contains no spatial derivatives). The time-like Carroll equation in the presence of such a potential is
\begin{align}
	\hbar^{2}\partial_{t}^{2}\psi(x,t) + U^{2}(t)\psi(x,t) + m_{0}^{2}\,\psi(x,t)=0.
\end{align}
We take a rectangular temporal barrier of height \(V\) and duration \(T = t_{2}-t_{1}\):
\begin{align}
	U(t) = 
	\begin{cases}
		0, & t < t_{1},\\
		V, & t_{1} \le t \le t_{2},\\
		0, & t > t_{2}.
	\end{cases}
\end{align}
Because the equation has no spatial derivatives, the spatial part \(\phi(x)\) factories and remains arbitrary; it cancels in the matching conditions. Hence we focus on the temporal factor \(\xi(t)\) with \(\psi(x,t)=\phi(x)\xi(t)\).

\paragraph{Region $I$ (pre-interaction, \(t<t_{1}\), \(U=0\)).}
The general solution is a superposition of the two frequency branches:
\begin{align}
	\psi_{I}(x,t)=\phi(x)\left[A e^{-i\frac{m_{0}}{\hbar}t} + B e^{i\frac{m_{0}}{\hbar}t}\right].
\end{align}
We choose the incoming state to be a pure positive frequency mode, so we set \(B=0\) and take \(A\) as the incident amplitude. This region establishes the reference state before the temporal barrier.

\paragraph{Region $II$ (inside the barrier, \(t_{1}\le t\le t_{2}\), \(U=V\)).}
Inside the barrier the effective frequency becomes \(\omega = \sqrt{m_{0}^{2}+V^{2}}\). The solution is
\begin{align}
	\psi_{II}(x,t)=\phi(x)\left[C e^{-i\frac{\omega}{\hbar}t} + D e^{i\frac{\omega}{\hbar}t}\right].
\end{align}
Even though the incident wave contains only a positive frequency \(m_{0}\), the discontinuity at \(t=t_{1}\) forces the appearance of both \(e^{-i\omega t}\) and \(e^{+i\omega t}\) components (i.e., \(D\neq0\)) because the field and its first time derivative must be continuous. This mode mixing is the temporal analogue of partial reflection and transmission at a spatial potential step.

\paragraph{Region $III$ (post-interaction, \(t>t_{2}\), \(U=0\)).}
After the barrier, the potential returns to zero, but the state is no longer pure:
\begin{align}
	\psi_{III}(x,t)=\phi(x)\left[E e^{-i\frac{m_{0}}{\hbar}t} + F e^{i\frac{m_{0}}{\hbar}t}\right].
\end{align}
The coefficient \(E\) represents the transmitted positive frequency part, while \(F\) is the amplitude of the generated negative frequency mode. The appearance of \(F\neq0\) is a direct signature of temporal scattering.

\paragraph{Physical interpretation.}
The temporal barrier acts as a non-adiabatic quench: the sudden change of the effective frequency from \(m_{0}\) to \(\omega\) and back to \(m_{0}\) mixes positive and negative frequency modes. This mixing is mathematically described by a Bogoliubov transformation.\footnote{A Bogoliubov transformation linearly relates initial and final mode amplitudes, mixing positive and negative frequency components while preserving the norm \(|E|^2-|F|^2=1\).} The conserved quantity is not the sum \(|E|^{2}+|F|^{2}\) but the Klein-Gordon norm \(|E|^{2}-|F|^{2}=1\), which reflects the indefinite inner product of the underlying field theory. Thus, the temporal barrier converts a fraction of the incident positive frequency mode into a negative frequency mode  a process that can be interpreted as “temporal reflection”. In the context of null surfaces, such frequency mixing is precisely the mechanism behind particle production by a time dependent background (e.g., a moving mirror or a collapsing shell). In the following subsection we will compute the coefficients \(E\) and \(F\) explicitly by matching the wavefunction and its time derivative at \(t=t_{1}\) and \(t=t_{2}\).

%----------------------------------------

%----------------------------------------
\subsubsection{Solutions}\label{subsubsec:temporal_tunneling_solutions}

We now solve explicitly the temporal scattering problem defined by the rectangular barrier \(U(t)\) (height \(V\), duration \(T = t_2 - t_1\)). Because the time-like Carroll equation contains no spatial derivatives, the spatial profile \(\phi(x)\) factorises and cancels in the matching conditions; hence we focus on the temporal factor. The general solution in each region is a superposition of positive and negative frequency exponentials:

\begin{align}
	\psi_{I}(x,t) &= \phi(x)\left[A\,e^{-i \frac{m_{0}}{\hbar} t} + B\,e^{i \frac{m_{0}}{\hbar} t}\right], && t < t_{1},\label{eq:sol_I}\\
	\psi_{II}(x,t) &= \phi(x)\left[C\,e^{-i \frac{\omega}{\hbar} t} + D\,e^{i \frac{\omega}{\hbar} t}\right], && t_{1} \le t \le t_{2},\label{eq:sol_II}\\
	\psi_{III}(x,t) &= \phi(x)\left[E\,e^{-i \frac{m_{0}}{\hbar} t} + F\,e^{i \frac{m_{0}}{\hbar} t}\right], && t_{2}<t,\label{eq:sol_III}
\end{align}
where
\[
\omega \equiv \sqrt{m_{0}^{2} + V^{2}}, \qquad T \equiv t_{2} - t_{1}.
\]

\paragraph{Incoming state and matching at \(t=t_{1}\).}
We consider a purely positive frequency incident mode from the left (Region $I$):
\[
\psi_{I}(x,t) = \phi(x)\,e^{-i \frac{m_{0}}{\hbar} t},
\]
which fixes \(A = 1,\; B = 0\). At the first interface \(t = t_{1}\) the wavefunction and its first time derivative must be continuous (since the equation is second order in time):
\begin{align}
	\psi_{I}(x,t_{1}) &= \psi_{II}(x,t_{1}), \\
	\partial_{t}\psi_{I}(x,t_{1}) &= \partial_{t}\psi_{II}(x,t_{1}).
\end{align}
Solving these two linear equations for \(C\) and \(D\) gives
\begin{align}
	C &= \frac{1}{2}\left(1 + \frac{m_{0}}{\omega}\right) e^{i(\omega - m_{0})t_{1}/\hbar},\\[4pt]
	D &= \frac{1}{2}\left(1 - \frac{m_{0}}{\omega}\right) e^{-i(\omega + m_{0})t_{1}/\hbar}.
\end{align}
These coefficients show that even though the incoming wave contains only a positive frequency \(m_{0}\), the sudden jump to \(\omega\) at \(t=t_{1}\) forces a non-zero negative frequency component \(D\) (unless \(V=0\)). This is the temporal analogue of partial reflection at a spatial potential step.

\paragraph{Matching at \(t=t_{2}\) and outgoing amplitudes.}
At the second interface \(t = t_{2}\) the potential returns to zero, so the effective frequency drops back to \(m_{0}\). Continuity of the wavefunction and its derivative at \(t=t_{2}\) yields
\begin{align}
	\psi_{II}(x,t_{2}) &= \psi_{III}(x,t_{2}), \\
	\partial_{t}\psi_{II}(x,t_{2}) &= \partial_{t}\psi_{III}(x,t_{2}).
\end{align}
Substituting the expressions for \(\psi_{II}\) and \(\psi_{III}\) and using the previously determined \(C\) and \(D\) leads to the outgoing amplitudes:
\begin{align}
	E &= \cos\!\left(\frac{\omega T}{\hbar}\right)
	- i\,\frac{2m_{0}^{2} + V^{2}}{2 m_{0}\omega}\,
	\sin\!\left(\frac{\omega T}{\hbar}\right), \\[6pt]
	F &= -\,i\,\frac{V^2}{2 m_{0}\omega}\,
	\sin\!\left(\frac{\omega T}{\hbar}\right).
\end{align}
These are the central results of temporal scattering.

The amplitude \(E\) (coefficient of \(e^{-i m_{0}t/\hbar}\) in Region $III$) represents the fraction of the incident positive frequency mode that is transmitted through the barrier without changing frequency sign. The amplitude \(F\) (coefficient of \(e^{+i m_{0}t/\hbar}\)) is the generated negative frequency component, which can be interpreted as a "temporal reflection". The squared magnitudes satisfy the Klein-Gordon norm conservation
\begin{align}
	|E|^{2}  = 1+|F|^{2}\,,
\end{align}
which is the time‑domain analogue of the unitarity relation \(|T|^{2}+|R|^{2}=1\) in ordinary Schrödinger scattering, but with a minus sign reflecting the indefinite inner product of the Carrollian theory.

In the limit \(V \to 0\) (no barrier), we have \(\omega = m_{0}\), and then \(E=1,\; F=0\) as expected. For finite \(V\), the amount of frequency mixing oscillates with the barrier duration \(T\) via \(\sin(\omega T/\hbar)\). This oscillatory behaviour is characteristic of coherent interference between the two components inside the barrier. The temporal tunneling phenomenon is therefore a genuine quantum effect with no spatial analogue in conventional quantum mechanics; it arises purely from the time-dependent potential and the degenerate causal structure of the Carrollian limit.
%-----------------------------------------------------------------------

%--------------------------------------------------------------------
\begin{figure}[h!]
\centering
\begin{tikzpicture}[scale=1.3]
\draw[->] (-0.5,0) -- (6,0) node[right] {$t$};
\draw[->] (0,-0.2) -- (0,3) node[left] {$U(t)$};

% Regions
\draw[thick] (0,0) -- (2,0);
\draw[thick] (2,1.5) -- (4,1.5);
\draw[thick] (4,0) -- (6,0);

% Vertical jumps
\draw[dashed] (2,0) -- (2,1.5);
\draw[dashed] (4,0) -- (4,1.5);

% Labels
\node at (1,1.3) {Region I};
\node at (3,2.2) {Region II};
\node at (5.2,1.3) {Region III};

\node at (2,-0.3) {$t_{1}$};
\node at (4,-0.3) {$t_{2}$};

\node at (3,1.7) {$V$};
\end{tikzpicture}
\caption{Temporal potential barrier $U(t)$ with height $V$ between $t_{1}$ and $t_{2}$.}
\end{figure}
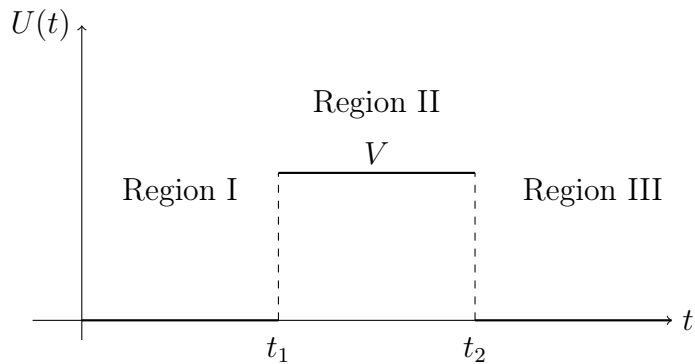
%--------------------------------------------------------------------

%---------------------------------------------------------------------
\subsubsection{Reflection and transmission coefficients}

We now analyse the scattering amplitudes that characterize the mixing of positive and negative frequencies induced by the temporal barrier. In Region $III$ the wavefunction is
\begin{align}
	\psi_{III}(x,t) = \phi(x)\!\left[E\,e^{-i \frac{m_{0}}{\hbar} t} + F\,e^{+i \frac{m_{0}}{\hbar} t}\right], \qquad t > t_{2},
\end{align}
where \(E\) and \(F\) are the outgoing positive and negative frequency amplitudes. Matching the solution across the temporal interface at \(t = t_{2}\) yields
\begin{align}
	E &= \cos\!\left(\frac{\omega T}{\hbar}\right) - i\,\frac{2m_{0}^{2}+V^{2}}{2 m_{0}\omega}\,\sin\!\left(\frac{\omega T}{\hbar}\right), \\[6pt]
	F &= -\,i\,\frac{V^2}{2 m_{0}\omega}\,\sin\!\left(\frac{\omega T}{\hbar}\right),
\end{align}
with
\[
\omega = \sqrt{m_{0}^{2} + V^{2}}, \qquad T = t_{2} - t_{1}.
\]

To make the frequency mixing more transparent, introduce the dimensionless parameters
\begin{align}
	\Gamma := \frac{V^2}{2 m_{0}\omega}, \qquad 
	\Delta := \frac{2m_{0}^{2}+V^{2}}{2 m_{0}\omega}, \qquad 
	\Theta := \frac{\omega T}{\hbar}.
\end{align}
The amplitudes then take the compact form
\begin{align}
	E = \cos\Theta - i\,\Delta \sin\Theta, \qquad
	F = -i\,\Gamma \sin\Theta.
\end{align}

\paragraph{Reflection and transmission amplitudes.}
A time dependent barrier mixes positive and negative frequency branches. The natural quantities characterizing this scattering are
\begin{align}
	\mathcal{R} := F, \qquad \mathcal{T} := E.
\end{align}
Their squared magnitudes quantify the strength of frequency conversion:
\begin{align}
	|\mathcal{R}|^{2} &= \Gamma^{2}\sin^{2}\Theta = \left(\frac{V^2}{2 m_{0}\omega}\right)^{2} \sin^{2}\!\left(\frac{\omega T}{\hbar}\right), \\[6pt]
	|\mathcal{T}|^{2} &= \cos^{2}\Theta + \Delta^{2}\sin^{2}\Theta 
	= \cos^{2}\!\left(\frac{\omega T}{\hbar}\right) + \left(\frac{2m_{0}^{2}+V^{2}}{2 m_{0}\omega}\right)^{2} \sin^{2}\!\left(\frac{\omega T}{\hbar}\right).
\end{align}
Thus \(\Gamma\) controls the generation of the negative frequency component (temporal reflection), while \(\Delta\) governs the distortion of the transmitted positive frequency mode.

Because spatial derivatives are absent in the Carroll limit, the wave equation reduces to an ensemble of decoupled time dependent oscillators. A "potential barrier" therefore cannot impede spatial propagation (Carroll particles cannot move); instead, it modifies the temporal evolution by changing the effective frequency. The scattering problem is intrinsically time-like, not spatial.

A temporal discontinuity forces the system to match \(\psi\) and \(\dot{\psi}\) across two time slices \(t=t_{1}\) and \(t=t_{2}\). Even if the incoming solution contains only positive frequency modes \(e^{-i m_{0}t/\hbar}\), these matching conditions inevitably generate a negative frequency component \(e^{+i m_{0}t/\hbar}\) after the barrier. This frequency mixing is the hallmark of a sudden frequency quench: the jump \(m_{0}\to\omega\) at \(t_{1}\) followed by \(\omega\to m_{0}\) at \(t_{2}\) transforms a pure mode into a linear combination of the two independent oscillator solutions.

The amplitudes \(E\) (positive frequency) and \(F\) (negative frequency) form a Bogoliubov transformation, preserving the Klein–Gordon norm. Hence
\[
|E|^{2}  = 1+|F|^{2}\,,
\]
just as spatial scattering obeys \(|T|^{2}+|R|^{2}=1\) in ordinary quantum mechanics. The temporal barrier thus acts as a \emph{temporal beam splitter}: part of the incoming mode is transmitted with the same frequency sign, the remainder is reflected into the opposite frequency sector. This highlights a distinctive feature of Carroll quantum mechanics: time dependent backgrounds can scatter a mode in its temporal orientation even though no spatial propagation is possible.

\paragraph{Klein-Gordon normalization.}
In a Klein-Gordon-type system, the inner product is indefinite: positive frequency modes carry positive norm, while negative frequency modes carry negative norm \cite{BjorkenDrell,GreinerRQM,SakuraiAQM}. The conserved quantity is not \(|\mathcal{T}|^{2}+|\mathcal{R}|^{2}\) but the Klein-Gordon norm
\[
|\mathcal{T}|^{2} = 1 + |\mathcal{R}|^{2}.
\]
The minus sign indicates that the reflected component is a negative frequency mode. Using the definitions above, one finds \(\Delta^{2}-\Gamma^{2}=1\), which expresses the unit determinant of the Bogoliubov transformation. Substituting the explicit amplitudes gives
\[
|E|^{2}-|F|^{2} = \cos^{2}\Theta + (\Delta^{2}-\Gamma^{2})\sin^{2}\Theta 
= \cos^{2}\Theta + \sin^{2}\Theta = 1.
\]

\paragraph{Physical reflection and transmission coefficients.}
Physically, \(|\mathcal{R}|^{2}\) measures the amount of negative frequency mode generated, while \(|\mathcal{T}|^{2}\) quantifies the surviving positive frequency component. We therefore define
\[
R_{\mathrm{temporal}} := |\mathcal{R}|^{2}, \qquad T_{\mathrm{temporal}} := |\mathcal{T}|^{2},
\]
as the temporal reflection and transmission coefficients. The relation \(T_{\mathrm{temporal}} - R_{\mathrm{temporal}} = 1\) (rather than \(T_{\mathrm{temporal}} + R_{\mathrm{temporal}} = 1\)) encodes the indefinite Klein-Gordon norm characteristic of Carrollian quantum dynamics, in contrast with the positive definite normalization of non-relativistic Schrödinger scattering.

\paragraph{Emergence of antiparticle-like excitations in the interacting theory.}
In the free time-like Carroll quantum mechanics, negative frequency modes are absent from the physical Hilbert space: they would correspond to negative rest energy and yield a negative probability density, and the Carroll group provides no charge conjugation symmetry to reinterpret them as antiparticles. However, when a time-dependent potential is present (temporal tunneling), the barrier dynamically mixes positive and negative frequencies. The outgoing state in Region $III$ contains a non-zero negative frequency component $F$ (the coefficient of $e^{+i m_0 t/\hbar}$). This generated negative frequency mode carries negative norm and can be interpreted as an \emph{antiparticle-like} excitation, although no fundamental antiparticle existed in the free theory. Thus, interactions in the time-like Carroll sector can create effective antiparticles, a phenomenon that mirrors the behavior of relativistic quantum fields where particle-antiparticle pairs are produced by time-dependent backgrounds. The relation $|\mathcal{T}|^2 - |\mathcal{R}|^2 = 1$ precisely encodes the indefinite norm that accommodates these emergent negative frequency states.

\subsubsection{Temporal Tunneling and Bogoliubov Processes}

Temporal tunneling in the time-like Carroll sector is closely related to familiar 
Bogoliubov transformations in relativistic quantum field theory.  
A time-dependent potential mixes positive and negative frequency modes, and the 
Carrollian wavefunction takes the form
\begin{align}
\phi(t) =
\begin{cases}
A\,e^{-i\omega_1 t} + B\,e^{+i\omega_1 t}, & t<0, \\[4pt]
C\,e^{-i\omega_2 t} + D\,e^{+i\omega_2 t}, & t>0,
\end{cases}
\end{align}
with matching at $t=0$ producing a Bogoliubov transformation
\begin{align}
\begin{pmatrix}
C \\ D
\end{pmatrix}
=
\begin{pmatrix}
\alpha & \beta \\
\beta^* & \alpha^*
\end{pmatrix}
\begin{pmatrix}
A \\ B
\end{pmatrix},
\qquad
|\alpha|^2 - |\beta|^2 = 1.
\end{align}

This structure mirrors particle creation in the dynamical Casimir effect, moving mirror 
models, and cosmological particle production, where non-stationary backgrounds induce 
frequency mixing.  
The key difference is that in the Carrollian limit \emph{spatial propagation is absent}, 
so the entire scattering process is encoded in the temporal phase structure.  
Temporal tunneling is therefore a purely time-domain analogue of Bogoliubov mixing, made 
exact and solvable by the ultra-relativistic contraction.
%--------------------------------------------------------------------
\subsection{Fracton/time-like Carroll QM Duality}\label{subsec:timelike_fracton}

Fracton phases of quantum matter host excitations with restricted mobility: isolated fractons are completely immobile, while bound states (dipoles) can move only along specific directions, giving rise to lineons and planons. These phases form a novel class of quantum states that bridges glassy dynamics, topological order, spin liquids, elasticity, quantum information, and gravity \cite{Chamon:2004lew,Bidussi:2021nmp,Pretko:2020cko,Gromov:2018nbv,Marsot:2022imf}. Recent work has examined connections between fractons and Carrollian particles \cite{Ahmadi-Jahmani:2025iqc,Figueroa-OFarrill:2023vbj,Marsot:2022imf,Figueroa-OFarrill:2023qty}. Here we develop a correspondence (or duality) between fractons and time‑like Carrollian quantum particles \cite{Figueroa-OFarrill:2023qty}. The correspondence rests on the classical particle models of \cite{Ahmadi-Jahmani:2025iqc}. We show that isolated fractons (monopoles) map to time‑like Carrollian QM particles, which are fundamentally immobile (no spatial derivatives in their quantum equations).

To verify the duality, we derive the quantum mechanical equation for a fractonic excitation. Consider a complex scalar field \(\phi(t,\mathbf{x})\) in \(d\) dimensions, with a Lagrangian that is invariant under \(U(1)\) and under dipole global and local symmetries. A minimal (lowest‑order) Lagrangian for such a theory can be written as
\begin{align}
	\mathcal{L} = |\partial_t\phi|^2 + m_0^2|\phi|^2 - \lambda\bigl|\phi\,\partial_i\partial_j\phi - \partial_i\phi\,\partial_j\phi - iA_{ij}\phi^2\bigr|^2,
\end{align}
where \(A_{ij}\) is a background tensor field that couples to the dipole moment, and repeated indices are summed. The coupling constant \(\lambda\) controls the strength of the dipole interaction; setting \(\lambda = 0\) turns off the interaction and leaves a free scalar field, which describes monopole fractons. 

Now reduce the theory to $0+1$ dimensions by assuming no spatial dependence 
(i.e., $\partial_i\phi = 0$). In this limit the Lagrangian simplifies to
\begin{equation}
    \mathcal{L}_{0+1} = |\dot\phi|^2 + m_0^2|\phi|^2,
    \label{eq:fracton_lagrangian_0+1}
\end{equation}
where the interaction term vanishes because it contains spatial derivatives.

A field theory in $0+1$ dimensions is equivalent to quantum mechanics: the field 
$\phi(t)$ becomes simply a time-dependent degree of freedom, and the Lagrangian 
\eqref{eq:fracton_lagrangian_0+1} describes a harmonic oscillator. For small 
excitations we may linearize the equations of motion. Restoring $\hbar$, the 
Euler--Lagrange equation for the linearized theory is
\begin{equation}
    \hbar^2\partial_t^2\phi + m_0^2\phi = 0,
    \label{eq:fracton_carroll_equation}
\end{equation}
which is exactly the time-like Carroll equation \eqref{time-like Carroll QM}. 
This shows that a fractonic monopole, when spatially homogeneous, behaves as a 
time-like Carroll particle, as previously noted in~\cite{Figueroa-OFarrill:2023qty}.

\paragraph{Isolated fracton / time‑like Carrollian QM particles.}
Time‑like Carroll QM particles with \(U(1)\) symmetry are dual to isolated fractons: both have no spatial derivatives in their quantum description. They are described by the wave equation
\[
(\hbar^{2}\partial_{t}^{2}+m_{0}^{2})\psi(\vec{x},t)=0,
\]
with \(\psi(\vec{x},t)\) a complex wavefunction. The absence of spatial derivatives reflects the complete immobility of these excitations, matching the fractonic property that an isolated fracton cannot move.

The duality established here provides a quantum‑mechanical realization of the Carroll/fracton correspondence. Isolated fractons (monopoles) are immobile by construction, and their quantum dynamics reduces to a simple harmonic oscillator in time – exactly the time‑like Carroll particle. The reduction to \(0+1\) dimensions (spatially uniform mode) is natural because a monopole has no internal spatial structure. This connection suggests that Carrollian quantum mechanics may serve as a tractable framework for studying fracton phases, especially their response to external fields and their entanglement properties.

%%%%%%%%%%%%%%%%%%%%%%%%%%%%%%%%%%%%%%%%%%%%%%%%%%%%%%%%%%%%%%%%%%%%%%%%%%%%%%%%%%%%%%%%%%%%%%%%%%%%%%%%%%%%%%%%%%%%%%%%%%%%%%%%%%%%%%%%%%%%%%%%%%%%%%%%%%%%%%%%%%%%%%%%%%%%%%%%%%%%%%%%%%%%%%%%%%%%%%%%%%%%%%%%%%%%%%%%%%%%
\section{Space-like Carroll QM}\label{sec:spacelike}

Unlike their time-like counterparts, space-like Carroll particles retain the ability to propagate spatially. However, invariance under Carroll boosts is not manifest in the naive formulation; to restore this symmetry, it is necessary to introduce compensating terms, which may either couple to the temporal derivative or appear as additive contributions independent of the temporal derivative \cite{Cotler:2024xhb}. In the coupled formulation, the energy spectrum becomes purely imaginary, indicating tachyonic behavior. Consequently, the probability density either identically vanishes or remains constant across all states. This leads to a non-standard quantum framework in which a vanishing density corresponds to zero-norm (null) states. While such states are generally considered unphysical in conventional quantum mechanics, they acquire physical significance within Carrollian physics, which is formulated on null geometries \footnote{The term "null" appears in two distinct but related contexts. Geometrically, a null surface is a hypersurface with a degenerate metric, characteristic of Carrollian spacetimes \cite{Ciambelli:2019lap,Ciambelli:2025unn}. Quantum mechanically, a null state is a state vector with zero norm, \(\langle \psi | \psi \rangle = 0\), which arises in the space-like sector from the vanishing density \(\rho = 0\). These two uses are not identical: one refers to the geometry of the spacetime, the other to the norm of the quantum state. However, they are intrinsically connected: the degenerate null geometry of the Carrollian spacetime is the fundamental reason why the quantum states acquire a zero norm. Thus, the same word "null" reflects the same underlying Carrollian structure manifesting at both the geometric and quantum levels.}. We have therefore demonstrated that, in contrast to the time-like sector—characterized by a positive density and a vanishing current—the space-like sector is distinguished by either vanishing or constant density accompanied by a non-zero spatial current.

The additive formulation, by contrast, reduces to a static constraint devoid of dynamics and a well-defined energy spectrum. These formulations correspond to fundamentally distinct physical regimes, underscoring the need for careful framework selection when modeling space-like Carrollian phenomena.

A key finding of this section is that classical space-like Carroll particles are tachyonic \cite{deBoer:2021jej,deBoer:2023fnj,Ahmadi-Jahmani:2025iqc,Bergshoeff:2022eog}. Their quantum analogues inherit this tachyonic nature, though its precise manifestation—especially in the energy spectrum, which is purely imaginary for both real and complex wavefunctions in the Hermitian formulation—depends on the nature of the wavefunction.

Recent studies have argued that Carrollian tachyons hold a meaningful physical role and warrant further investigation \cite{Ecker:2024czx}. Moreover, in light of the established framework of Carroll black hole thermodynamics \cite{Ecker:2023uwm}, it is plausible to anticipate the existence of scalar modes that function as carriers of a Carrollian analogue of Hawking radiation. Additionally, theoretical actions describing Carrollian tachyonic particles have been developed and analyzed \cite{Bergshoeff:2022qkx,Casalbuoni:2023bbh}, further underscoring the physical viability and utility of these excitations in the Carrollian regime.

%-----------------------------------------%
\subsection{Derivation of the space-like Carroll equation}

The space-like Carrollian quantum equation admits two distinct derivations, each yielding different mathematical structures and physical interpretations. The first approach couples a temporal derivative to the compensating field \(\chi\). It is characterized by Carroll invariance, linearity, a well-defined energy spectrum, and a consistent probabilistic interpretation, providing a genuine quantum mechanical description of tachyonic Carrollian particles. The second approach yields an equation devoid of explicit temporal derivatives, ensuring Carroll invariance through the compensating field \(\tilde{\chi}\). This formulation presents several fundamental challenges: it constitutes a constraint rather than an evolution equation, lacks a well-defined energy spectrum, and admits an ambiguous physical interpretation. Consequently, while it does not provide a standard quantum mechanical framework with explicit time evolution and probabilistic interpretation, it may describe systems where time dependence is encoded implicitly through the background field \(\tilde{\chi}\).

In what follows, we examine these two approaches in detail, beginning with the first, which forms the basis of our quantum mechanical treatment.

\paragraph{Approach 1:}
The derivation branches into two cases depending on whether the wavefunction is real or complex. For real \(\psi\), the compensating field \(\chi\) is real, and the energy is purely imaginary, signaling tachyonic behavior while preserving the Born rule, though without the standard interpretation as a conserved probability. For complex \(\psi\), Hermiticity requires \(\chi\) to be purely imaginary, yielding a consistent probabilistic framework with a modified density. Both cases are fully Carroll-invariant, linear, and admit well-defined energy spectra, as discussed below.

\paragraph{Real wavefunction.}
Applying the space-like Carroll limit \eqref{space-like Carroll limit} to the Klein-Gordon equation \eqref{Klein-Gordon Equation} for a real wavefunction gives
\begin{align}
    \bigl(-\hbar^{2}\partial_{x}^{2}+m_{0}^{2}\bigr)\psi(\vec{x},t)=0,
\end{align}
which is not Carroll boost invariant. To restore invariance, a compensating field \(\chi\) is introduced, leading to the modified wave equation
\begin{align}\label{space-like Carroll QM}
    \bigl(\hbar^{2}\chi\,\partial_{t}-\hbar^{2}\partial_{x}^{2}+m_{0}^{2}\bigr)\psi(\vec{x},t)=0.
\end{align}
Under a Carroll boost with parameter \(\vec{\beta}\), the field \(\chi\) transforms as
\begin{align}
    \delta_{C}\chi = \beta^{2}\partial_{t} + 2\,\vec{\beta}\cdot\vec{\partial}_{x},
\end{align}
while \(\psi\) transforms as a scalar: \(\psi(\vec{x},t)=\psi'(\vec{x}',t')\). The presence of \(\chi\) is essential: it is a non-dynamical background field that compensates for the non-invariance of the spatial Laplacian, thereby restoring boost invariance.

\paragraph{Complex wavefunction.}
When the wavefunction is complex, the presence of \(U(1)\) symmetry requires the explicit imaginary unit \(i\) in the equation to maintain Hermiticity:
\begin{align}
    (i\hbar^{2}\chi\,\partial_{t} - \hbar^{2}\partial_{x}^{2} + m_{0}^{2})\psi = 0, \qquad \psi \in \mathbb{C}. \label{eq:spacelike_complex}
\end{align}
In this case, \(\chi\) must be purely imaginary, and its transformation under Carroll boosts is
\begin{align}
    \delta_{C}\chi = -i\beta^{2}\partial_{t} - 2i\,\vec{\beta}\cdot\vec{\partial}_{x}. \label{eq:boost_transform}
\end{align}
This transformation is necessary to cancel the extra terms arising from the spatial Laplacian while preserving the Hermitian structure.

The complex wavefunction sector is fundamentally tachyonic, as reflected in the purely imaginary energy spectrum. We also demonstrate that, despite this tachyonic nature, probability conservation holds for energy eigenstates in the Hermitian formulation, as will be shown in the following discussion.

\paragraph{Approach 2:}
An alternative derivation yields the constraint equation
\begin{align}
    \hbar^{2}\tilde{\chi} + \left(- \hbar^{2}\partial_{x}^{2} + m_{0}^{2}\right) \psi = 0. \label{second Space-like eq}
\end{align}
Here, the field \(\tilde{\chi}\) is not coupled to \(\psi\); it enters as an additive term. Carroll invariance is achieved through the transformation law
\begin{align}
    \delta_C \tilde{\chi} = 2\vec{\beta}\cdot\partial_{x}\partial_t\psi + \beta^2\partial_t^2\psi. 
\end{align}
This equation is linear in \(\psi\) and \(\tilde{\chi}\), but it contains no explicit time derivatives. The field \(\tilde{\chi}\) is determined entirely by \(\psi\) and its spatial derivatives via
\begin{align}
    \hbar^{2}\tilde{\chi} = \hbar^{2}\partial_{x}^{2}\psi - m_{0}^{2}\psi.
\end{align}
While this approach may be relevant for equilibrium configurations or gauge-fixed formulations, it does not provide a viable quantum mechanical framework, as it lacks a well-defined energy spectrum, explicit time evolution, and a probabilistic interpretation.

We therefore adopt Approach 1 as the physically meaningful and mathematically consistent framework for space-like Carrollian quantum mechanics.

\subsection{The compensating fields \texorpdfstring{$\chi$}{chi} and \texorpdfstring{$\tilde{\chi}$}{chi}: roles and interpretation}

The field $\chi$ in \eqref{space-like Carroll QM} is a non-dynamical background field introduced to restore Carroll boost invariance \cite{Bergshoeff:2022qkx}. It can be regarded as a background structure; for further details, see \cite{Bruce:2026knl}. Without it, the naive equation $(-\hbar^2\partial_{x}^2 + m_0^2)\psi = 0$ is not invariant under Carroll boosts $t' = t - \vec{\beta}\cdot\vec{x}$. The invariant equation requires $\chi$ to transform as a compensating field, canceling the non-invariance of the spatial Laplacian. Its explicit transformation law is not needed for the present analysis.

Although its transformation involves derivatives, $\chi$ is a classical field, not a quantum operator. It has no kinetic term and is not quantized; it acts as a St\"uckelberg-like connection. The derivatives in its transformation are part of the classical gauge structure, rather than quantum operators acting on a Hilbert space.

A crucial observation is that the nature of $\chi$ and its transformation depend on whether the wavefunction is real or complex. For a real wavefunction, $\chi$ is real and transforms as $\delta_C\chi = \beta^2\partial_t + 2\vec{\beta}\cdot\vec{\partial}_{x}$. For a complex wavefunction, $\chi$ is complex and transforms with an explicit $i$. This distinction has profound implications for the Hermiticity of the equation and the physical interpretation of the space-like Carroll sector.

The field $\tilde{\chi}$ in \eqref{second Space-like eq} plays a fundamentally different role. Unlike $\chi$, which is an independent compensating field, $\tilde{\chi}$ is not an independent field but a functional of $\psi$, encoding its spatial structure:
\begin{align}
    \tilde{\chi} = \partial_{x}^{2}\psi - \frac{m_{0}^{2}}{\hbar^{2}}\psi. 
\end{align}
It possesses no independent dynamics of its own; rather, it is entirely determined by $\psi$ and its spatial derivatives. While Approach 2 is Carroll-invariant, it does not provide a viable quantum mechanical framework, as it lacks an explicit time evolution, a well-defined energy spectrum, and a probabilistic interpretation. However, it may find utility in equilibrium configurations or field-theoretic extensions.

In contrast to Approach 1, where $\chi$ is an independent compensating field coupled to \(\partial_t\psi\), Approach 2 features \(\tilde{\chi}\) as a dependent field determined by \(\psi\). This distinction reflects fundamentally different physical regimes: Approach 1 provides a genuine quantum mechanical description of tachyonic Carrollian particles, while Approach 2 describes a constraint-based system with dynamics encoded through the background field rather than explicit time derivatives.
%%----------------------------------------%%

\subsection{Probability conservation and continuity equation}

For charged (or $U(1)$-symmetric) spacelike Carroll particles, the wavefunction is complex. The conjugate equation reads
\begin{align}
    (-i\hbar^{2}\chi^*\,\partial_{t} - \hbar^{2}\partial_{x}^{2} + m_{0}^{2})\psi^* &= 0.
    \label{eq:spacelike_complex_conj}
\end{align}
The system is invariant under Carroll boost transformations provided $\chi$ and $\chi^*$ transform appropriately (see Eq.~\eqref{eq:boost_transform}). Following the standard procedure, we multiply Eq.~\eqref{eq:spacelike_complex} by $\psi^*$ and Eq.~\eqref{eq:spacelike_complex_conj} by $\psi$, then subtract the second from the first. This yields the intermediate identity
\begin{align}
    i\hbar^{2}\left(\chi\,\psi^*\partial_{t}\psi + \chi^*\,\psi\partial_{t}\psi^*\right)
    -\hbar^{2}\left(\psi^*\partial_{x}^{2}\psi - \psi\partial_{x}^{2}\psi^*\right) = 0.
    \label{eq:pre_continuity}
\end{align}
Since $\chi$ is purely imaginary, we parametrise $\chi = i\chi'$ with $\chi' \in \mathbb{R}$. Substituting this into Eq.~\eqref{eq:pre_continuity} gives
\begin{align}
    -\hbar^{2}\,\chi'\,\left(\psi^*\partial_{t}\psi - \psi\partial_{t}\psi^*\right)
    -\hbar^{2}\,\partial_x\left(\psi^*\partial_x\psi - \psi\partial_x\psi^*\right) = 0.
\end{align}
Multiplying by $-1$, we obtain the continuity equation
\begin{align}
    \partial_t \rho + \partial_x J = 0,
    \label{eq:continuity_carroll}
\end{align}
where we identify the current and the time derivative of the density as
\begin{align}
    J &= \hbar^{2}\left(\psi^*\partial_x\psi - \psi\partial_x\psi^*\right), \label{eq:J_carroll} \\[4pt]
    \rho &= \hbar^{2}\chi' \int^t dt' \, \left(\psi^*\partial_{t}\psi - \psi\partial_{t}\psi^*\right). \label{eq:rho_carroll}
\end{align}
The current $J$ defined in Eq.~\eqref{eq:J_carroll} is generally imaginary for propagating modes. This is not a problem mathematically, as the continuity equation holds regardless of whether $J$ is real or complex. However, for physical interpretation, it is convenient to define the \emph{physical flux}
\begin{align}
    \mathcal{J} \equiv -i J = -i\hbar^{2}\left(\psi^*\partial_x\psi - \psi\partial_x\psi^*\right).
    \label{eq:physical_flux}
\end{align}
The density itself is therefore defined up to an additive constant:
\begin{align}
     \partial_t \rho &= \hbar^{2}\chi' \left(\psi^*\partial_t\psi - \psi\partial_t\psi^*\right). \label{eq:rho_dot_carroll}
\end{align}
For energy eigenstates, we impose $E\psi = i\hbar\,\partial_t \psi$. The space-like (tachyonic) dispersion relation yields purely imaginary energies, $E = \pm i\Gamma$ with $\Gamma \in \mathbb{R}$, so that $E = -E^*$. Substituting this into Eq.~\eqref{eq:rho_dot_carroll}, we find
\begin{align}
    \psi^*\partial_t\psi - \psi\partial_t\psi^*
    = -\frac{i(E+E^*)}{\hbar}|\psi|^2 = 0\,,
\end{align}
consequently,
\begin{align}
    \partial_t \rho = 0, \qquad \rho = C \quad (\text{constant}),
\end{align}
where $C$ is an integration constant. For the pure null sector, we fix the gauge by imposing $\rho \to 0$ at spatial infinity, yielding $C = 0$. Thus, for space-like modes, the local density is either a uniform background ($C \neq 0$) or vanishes identically ($C = 0$).

The space-like Carroll sector admits spatial propagation only under a strict geometric constraint. In Carrollian geometry, the causal structure is degenerate and ordinarily forbids motion; propagation can occur only along the null direction of the Carrollian metric. This requirement is encoded in the continuity equation, which forces the probability density to vanish for propagating states, $\rho = 0$. Thus, the condition $\rho = 0$ is not merely a gauge choice but a physical constraint arising from the underlying Carrollian geometry. States in this sector are therefore strictly null: they possess zero norm, and their physical content resides entirely in the non-vanishing spatial flux
\[
\mathcal{J} = -i\hbar^{2}\left(\psi^{*}\partial_{x}\psi - \psi\,\partial_{x}\psi^{*}\right).
\]
In contrast to standard quantum mechanics, where zero-norm states are typically discarded as unphysical, the Carrollian framework embraces them as genuine degrees of freedom. Observable quantities are not localized probabilities but fluxes propagating along the null direction of Carrollian spacetime. The spacelike Carroll sector is therefore a pure null sector, with $\rho = 0$ emerging as the natural and physically required condition for spatial propagation.
%%%---------------------------------------%%
\subsection{Space-like Carrollian particles in a box}

We proceed to solve the space-like Carroll equation for a particle confined 
within a one-dimensional infinite potential well. We first examine the 
complex wavefunction case, followed by the real wavefunction case.

The Hermitian space-like Carroll equation for a complex wavefunction is:
\begin{align*}
    \left(i\hbar^{2}\chi\,\partial_{t}
    -\hbar^{2}\partial_{x}^{2}
    +m_{0}^{2}\right)\psi(x,t)=0\,.
\end{align*}
The infinite potential well is characterized by a spatial potential that 
vanishes within the interval $0 < x < L$ and is infinite outside this 
region, thereby enforcing the boundary conditions 
$\psi(0,t) = \psi(L,t) = 0$.

\paragraph{Separation of variables.}
Employing the method of separation of variables, we decompose the 
wavefunction as
\begin{align}
    \psi(x,t) = \phi(x)\,\xi(t).
\end{align}
Substituting this ansatz into the governing equation and dividing by the 
product $\phi(x)\xi(t)$ yields
\begin{align}
    i\hbar^{2}\chi\,\frac{\dot{\xi}}{\xi} - \frac{\hbar^{2}\phi''}{\phi} = -m_{0}^{2}.
\end{align}
Since the left-hand side is the sum of a function of $t$ only and a 
function of $x$ only, each term must independently equal a constant. 
To obtain oscillatory spatial solutions, we adopt the following sign 
convention:
\begin{align}
    i\hbar^{2}\chi\,\dot{\xi}(t) &= -\lambda^{2}\,\xi(t), \label{eq:SL_time_final}\\
    -\hbar^{2}\phi''(x) &= (\lambda^{2} - m_{0}^{2})\phi(x). \label{eq:SL_space_final}
\end{align}
This choice of sign is essential for the particle-in-a-box problem, as 
it produces sinusoidal spatial solutions $\phi(x) \propto \sin(kx)$ 
that satisfy the boundary conditions $\phi(0)=\phi(L)=0$.

\paragraph{Temporal part.}
Equation \eqref{eq:SL_time_final} can be rearranged as
\begin{align}
    \dot{\xi}(t) = i\frac{\lambda^{2}}{\hbar^{2}\chi}\,\xi(t),
\end{align}
which is a first-order linear differential equation. Its general solution is
\begin{align}
    \xi(t) = A\,\exp\!\left[i\frac{\lambda^{2}}{\hbar^{2}\chi}\,t\right],
\end{align}
where $A$ is a complex integration constant.

Comparing this with the standard quantum mechanical phase factor 
$e^{-iEt/\hbar}$, we identify the energy eigenvalue as
\begin{align}
    E = -\frac{\lambda^2}{\hbar \chi}.
    \label{eq:energy_sign_conv}
\end{align}
For the Hermitian case, where $\chi = i\chi'$ with $\chi' \in \mathbb{R}$, this becomes purely imaginary:
\begin{align}
    E = i\frac{\lambda^{2}}{\hbar \chi'}.
    \label{eq:energy_imaginary}
\end{align}
This purely imaginary spectrum reflects the tachyonic nature of the complex space-like sector.

\paragraph{Spatial part.}
Equation \eqref{eq:SL_space_final} can be rewritten as:
\begin{align}
    \frac{d^{2}\phi}{dx^{2}}(x) + k^{2}\phi(x) = 0,
    \quad \text{where} \quad
    k^{2} = \frac{\lambda^{2} - m_{0}^{2}}{\hbar^{2}}.
\end{align}
The general solution is:
\begin{align}
    \phi(x) = B \sin(kx) + C \cos(kx).
\end{align}
The boundary conditions $\phi(0) = \phi(L) = 0$ give $C = 0$ and 
$\sin(kL) = 0$, leading to:
\begin{align}
    kL = n\pi \quad \Rightarrow \quad k_n = \frac{n\pi}{L}, \qquad n = 1,2,3,\dots
\end{align}
Substituting this into the definition of $k^2$ gives:
\begin{align}
    \frac{\lambda_{n}^{2} - m_{0}^{2}}{\hbar^{2}} = \left(\frac{n\pi}{L}\right)^{2}
    \quad \Rightarrow \quad
    \lambda_{n}^{2} = \hbar^{2}\left(\frac{n\pi}{L}\right)^{2} + m_{0}^{2}.
\end{align}

The spatial eigenfunctions are:
\begin{align}
    \phi_{n}(x) = B_{n} \sin\left(\frac{n\pi x}{L}\right).
\end{align}
The full time-dependent wavefunction is:
\begin{align}
    \psi_{n}(x,t) = B_{n} \sin\left(\frac{n\pi x}{L}\right) 
    \exp\!\left[i \frac{\lambda_{n}^{2}}{\hbar^{2}\chi} t \right],
\end{align}
where
\begin{align}
    \lambda_{n}^{2} = \hbar^{2}\left(\frac{n\pi}{L}\right)^{2} + m_{0}^{2}.
\end{align}

Using the sign convention in \eqref{eq:energy_sign_conv}, the energy 
eigenvalues are:
\begin{align}
    E_{n} = -\frac{\lambda_{n}^{2}}{\hbar \chi}
    = -\frac{\hbar}{\chi}\left(\frac{n\pi}{L}\right)^{2} - \frac{m_{0}^{2}}{\hbar \chi}.
    \label{eq:energy_spectrum_sign}
\end{align}
For the Hermitian case, with $\chi = i\chi'$, this becomes:
\begin{align}
    E_{n} = i\frac{\lambda_{n}^{2}}{\hbar \chi'}
    = i\frac{\hbar}{\chi'}\left(\frac{n\pi}{L}\right)^{2} + i\frac{m_{0}^{2}}{\hbar \chi'}.
    \label{eq:energy_spectrum_imaginary}
\end{align}
This purely imaginary spectrum reflects the tachyonic nature of the complex space-like sector.

%%----------------------------------------%%
\subsection{Space-like tunneling}

As discussed in the previous sections, tunneling phenomena can indeed arise 
within Carrollian quantum dynamics. In this subsection, we provide a 
comprehensive analysis of the tunneling behavior of space-like Carrollian 
quantum particles. The dynamics of these excitations are governed by the 
space-like Carrollian equation of motion in the presence of a potential:
\begin{align}
    \bigl(i\hbar^{2}\chi\,\partial_{t}-\hbar^{2}\partial_{x}^{2}+U^{2}(x,t)+m_{0}^{2}\bigr)\psi(x,t)=0.
\end{align}
The potential barrier may possess both spatial and temporal structure. We 
consider a static rectangular barrier of height $V_0$ and width $L = x_2 - x_1$:
\begin{align}
    U(x) = 
    \begin{cases}
        0, & x < x_1,\\
        V_0, & x_1 \le x \le x_2,\\
        0, & x > x_2.
    \end{cases}
\end{align}
Because the potential is time-independent, we can 
separate variables: \(\psi(x,t)=\phi(x)\xi(t)\). Substituting into the 
equation and dividing by \(\phi(x)\xi(t)\) gives
\begin{align}
    i\hbar^{2}\chi\,\frac{\dot{\xi}(t)}{\xi(t)} - \frac{\hbar^{2}\phi''(x)}{\phi(x)} + U^{2}(x) + m_{0}^{2} = 0.
\end{align}
The first term depends only on \(t\); the rest depend only on \(x\). Hence they 
must equal a constant, which we denote by \(-\lambda^{2}\). This yields
\begin{align}
    i\hbar^{2}\chi\,\dot{\xi}(t) &= -\lambda^{2}\,\xi(t), \label{eq:time_tunnel}\\
    -\hbar^{2}\phi''(x) + \bigl(U^{2}(x)+m_{0}^{2}\bigr)\phi(x) &= \lambda^{2}\,\phi(x). \label{eq:space_tunnel}
\end{align}

\paragraph{Temporal part.}
Equation \eqref{eq:time_tunnel} gives
\begin{align}
    \xi(t) = \xi_{0}\,e^{i\frac{\lambda^{2}}{\hbar^{2}\chi}\,t}.
\end{align}
The temporal evolution is oscillatory, with frequency \(\omega = \lambda^2/(\hbar^2\chi)\).

\paragraph{Spatial part.}
In each region where \(U(x)\) is constant, equation \eqref{eq:space_tunnel} 
becomes
\begin{align}
    -\hbar^{2}\phi''(x) + (U^{2}+m_{0}^{2})\phi(x) = \lambda^{2}\,\phi(x).
\end{align}
Rearranging,
\begin{align}
    \phi''(x) + \frac{\lambda^{2} - U^{2} - m_{0}^{2}}{\hbar^{2}}\,\phi(x)=0.
\end{align}

\paragraph{Region I (\(x<x_{1}\), \(U=0\)).}
Here the spatial equation is \(\phi'' + k^{2}\phi=0\) with
\begin{align}
    k^{2} = \frac{\lambda^{2} - m_{0}^{2}}{\hbar^{2}}.
\end{align}
The general solution is a superposition of right and left moving plane waves:
\begin{align}
    \phi_{I}(x) = A e^{ikx} + B e^{-ikx}.
\end{align}

\paragraph{Region II (\(x_{1}\le x\le x_{2}\), \(U=V_0\)).}
The equation becomes \(\phi'' - \kappa^{2}\phi=0\) with
\begin{align}
    \kappa^{2} = \frac{V_0^{2} + m_{0}^{2} - \lambda^{2}}{\hbar^{2}}.
\end{align}
For a barrier we require \(\kappa^{2}>0\), so the solutions are exponential:
\begin{align}
    \phi_{II}(x) = C e^{\kappa x} + D e^{-\kappa x}.
\end{align}

\paragraph{Region III (\(x>x_{2}\), \(U=0\)).}
This region is identical to Region I, with the same wave number \(k\). We 
assume no incoming wave from the right, so the solution is
\begin{align}
    \phi_{III}(x) = F e^{ikx} + G e^{-ikx},\qquad G=0.
\end{align}

\paragraph{Matching conditions.}
Continuity of \(\phi\) and \(\phi'\) at \(x=x_{1}\) and \(x=x_{2}\) gives 
four equations. The common temporal factor \(\xi(t)\) cancels, so we work 
solely with \(\phi(x)\). Explicitly:
\begin{align}
    A e^{ikx_{1}} + B e^{-ikx_{1}} &= C e^{\kappa x_{1}} + D e^{-\kappa x_{1}},\\
    ik\big(A e^{ikx_{1}} - B e^{-ikx_{1}}\big) &= \kappa\big(C e^{\kappa x_{1}} - D e^{-\kappa x_{1}}\big),\\[2mm]
    C e^{\kappa x_{2}} + D e^{-\kappa x_{2}} &= F e^{ikx_{2}},\\
    \kappa\big(C e^{\kappa x_{2}} - D e^{-\kappa x_{2}}\big) &= ik\,F e^{ikx_{2}}.
\end{align}
We set \(A\) as the incident amplitude, \(B\) the reflected amplitude, and 
\(F\) the transmitted amplitude. The barrier width is \(L=x_{2}-x_{1}\). 
It is convenient to shift the origin to \(x_{1}\) by setting \(y=x-x_{1}\); 
then the barrier occupies \(0\le y\le L\).

\paragraph{Solution for the coefficients.}
Solving the linear system yields the coefficients inside the barrier:
\begin{align}
    C &= \frac{1}{2}\left[\left(1+\frac{ik}{\kappa}\right)A + \left(1-\frac{ik}{\kappa}\right)B\right] e^{-\kappa x_{1}},\\[4pt]
    D &= \frac{1}{2}\left[\left(1-\frac{ik}{\kappa}\right)A + \left(1+\frac{ik}{\kappa}\right)B\right] e^{\kappa x_{1}}.
\end{align}
The transmission and reflection amplitudes are
\begin{align}
    \mathcal{T} &\equiv \frac{F}{A} = \frac{4k\kappa\,e^{-ikL}}{(k+\kappa)^{2}e^{\kappa L} - (k-\kappa)^{2}e^{-\kappa L}},\\[4pt]
    \mathcal{R} &\equiv \frac{B}{A} = \frac{(k^{2}-\kappa^{2})\sinh(\kappa L)}{2ik\kappa\cosh(\kappa L) + (k^{2}+\kappa^{2})\sinh(\kappa L)}.
\end{align}
The transmission probability is \(T = |\mathcal{T}|^{2}\) and the reflection 
probability \(R = |\mathcal{R}|^{2}\); they satisfy \(T+R=1\) as expected for 
a Hermitian problem.

%-----------------------------------------------------------------------
%-----------------------------------------------------------------------
%-----------------------------------------------------------------------
%-----------------------------------------------------------------------

%-----------------------------------------------------------------------
%-----------------------------------------------------------------------
%-----------------------------------------------------------------------

%%%%%%%%%%%%%%%%%%%%%%%%%%%%%%%%%%%%%%%%%%%%%%%%%%%%%%%%%%%%%%%%%%%%%%%%%%%%%%%%%%%%%%%%%%%%%%%%%%%%%%%%%%%%%%%%%%%%%%%%%%%%%%%%%%%%%%%%%%%%%%%%%%%%%%%%%%%%%%%%%%%%%%%%%%%%%%%%%%%%%%%%%%%%%%%%%%%%%%%%%%%%%%%%%%%%%%%%%%%%%%%%%%%%%%%%%%%%%%%%%%%%%%%%%%%%%%%%%%%%%%%%%%%%%%%%%%%%%%%%%%%%%%%%%%%%%%%%%%%%%%%%%%%%%%%%%%%%%%%%%%%%%%%%%%%%%%%%%%%%%%%%%%%%%%%%%%%%%%%%%%%%%%%%%%%%%%%%%%%%%%%%%%%%%%%%%%%%%%%%%%%%%%%%%%%%%%%%%%%%%%%%%%%%%%
\section{Hybrid Carrollian QM}\label{sec:hybrid}

Having separately analyzed the time-like and space-like sectors, we now explore a unified Carrollian framework that retains both temporal and spatial derivatives. This \textbf{hybrid} sector arises from a specific contraction \eqref{hybrid Carroll limit} of the Klein-Gordon equation \eqref{Klein-Gordon Equation}, preserving both \(\partial_t^2\) and \(\partial_x^2\). As in the space-like sector, restoring Carroll boost invariance requires the introduction of a compensating field. This field may either couple explicitly to the temporal derivative (Approach 1) or appear as an additive term in a constraint equation devoid of such derivatives (Approach 2).

An alternative route to the hybrid Carrollian sector, independent of the limiting procedure, consists of regarding the Klein-Gordon equation as the seed equation. Carroll boost invariance is subsequently restored through the introduction of a compensating term, which may couple to the temporal derivative or appear as an additive contribution in a constraint equation.

In the hybrid Carroll sector, tachyonic behavior emerges in Approach 1---where the first-order temporal derivative couples to the compensating field \(\chi\)---provided the wavefunction is complex. In this case, complex energy eigenvalues signal tachyonic behavior. When the wavefunction is real, however, the equation reduces to a damped harmonic oscillator with real coefficients, and the sector is non-tachyonic. By contrast, Approach 2, which is devoid of a temporal derivative, reduces to an inhomogeneous Klein-Gordon equation and does not exhibit tachyonic behavior. The presence or absence of the coupled temporal derivative thus distinguishes between tachyonic and non-tachyonic regimes in the hybrid sector.

The hybrid sector occupies an intermediate position between the time-like and space-like limits. Unlike the space-like sector, it is not a null sector: the presence of both first-order and second-order time derivatives allows for mixed dynamics, and generically \(\rho \neq 0\). Only in special limits—when the first-order term dominates—does it reduce to the pure null behavior of the space-like sector. Thus, the hybrid sector contains a null subspace as a limiting case, while supporting non-null states with non-vanishing density.

%%----------------------------------------%%
\subsection{Derivation of the hybrid wave equation}

We now derive the hybrid Carrollian quantum equation via the contraction procedure, following the two distinct approaches identified in the space-like sector: one in which the compensating field couples explicitly to the temporal derivative (Approach 1), and another in which it appears as an additive term in a constraint equation devoid of such derivatives (Approach 2).

\paragraph{Approach 1:}
In this approach, the first-order temporal derivative term \(\hbar^{2}\chi\,\partial_{t}\psi\) emerges as the compensating contribution required to restore Carroll boost invariance. The nature of the compensating field \(\chi\) is determined by the wavefunction: for real \(\psi\), \(\chi\) is real; for complex \(\psi\), Hermiticity requires \(\chi\) to be purely imaginary.

\paragraph{Real wavefunction.}
Applying the hybrid contraction limit \eqref{hybrid Carroll limit} to the Klein-Gordon equation \eqref{Klein-Gordon Equation} and requiring Carroll boost invariance yields the hybrid Carrollian wave equation
\begin{align}\label{Hybrid Carroll QM}
	\bigl(\hbar^{2}\partial_{t}^{2} - \hbar^{2}\partial_{x}^{2} + \hbar^{2}\chi\,\partial_{t} + m_{0}^{2}\bigr)\psi(\vec{x},t) = 0.
\end{align}
Here, \(\chi\) is the same compensating field introduced in the space-like sector; its presence ensures invariance under Carroll boosts. The equation contains both second-order time derivatives (characteristic of the time-like sector) and a first-order time derivative (characteristic of the space-like sector), alongside spatial Laplacian terms. Thus, the hybrid sector genuinely mixes the properties of the two previous sectors.

\paragraph{Complex wavefunction.}
When the wavefunction is complex, the presence of \(U(1)\) symmetry requires the explicit imaginary unit \(i\) in the hybrid equation to maintain Hermiticity:
\begin{align}
    \left(\hbar^{2}\partial_{t}^{2} - \hbar^{2}\partial_{x}^{2} + i\hbar^{2}\chi\,\partial_{t} + m_{0}^{2}\right)\psi = 0, \qquad \psi \in \mathbb{C}. \label{eq:hybrid_complex}
\end{align}
In this case, \(\chi\) must be purely imaginary, so we write \(\chi = i\chi'\) with \(\chi' \in \mathbb{R}\). Substituting this into the equation yields
\begin{align}
    \left(\hbar^{2}\partial_{t}^{2} - \hbar^{2}\partial_{x}^{2} - \hbar^{2}\chi'\partial_{t} + m_{0}^{2}\right)\psi = 0.\label{eq:hybrid_complex1}
\end{align}
The transformation of \(\chi'\) under Carroll boosts is
\begin{align}
    \delta_{C}\chi' = -\beta^{2}\partial_{t} - 2\,\vec{\beta}\cdot\vec{\partial}_{x}.
\end{align}
This transformation is necessary to cancel the extra terms arising from the spatial Laplacian while preserving the Hermitian structure. The complex wavefunction sector of the hybrid Carrollian theory is fundamentally tachyonic.

\paragraph{Approach 2:}
An alternative derivation yields the constraint equation
\begin{align}
    \left(\hbar^{2}\partial_{t}^{2} - \hbar^{2}\partial_{x}^{2} + m_{0}^{2}\right)\psi + \hbar^{2}\tilde{\chi} = 0. \label{hybrid_approach2}
\end{align}
Here, the field \(\tilde{\chi}\) is not coupled to \(\psi\); it enters as an additive term. Carroll invariance is achieved through the transformation law
\begin{align}
    \delta_C \tilde{\chi} = 2\vec{\beta}\cdot\partial_{x}\partial_t\psi + \beta^2\partial_t^2\psi. 
\end{align}
This equation is linear in \(\psi\) and \(\tilde{\chi}\), but it contains no explicit first-order temporal derivative coupled to \(\psi\). The field \(\tilde{\chi}\) is determined entirely by \(\psi\) and its derivatives via
\begin{align}
    \hbar^{2}\tilde{\chi} = -\hbar^{2}\partial_{t}^{2}\psi + \hbar^{2}\partial_{x}^{2}\psi - m_{0}^{2}\psi.
\end{align}
The energy spectrum is obtained from the homogeneous part and is purely real:
\begin{align}
    E = \pm \sqrt{\hbar^{2}k^{2} + m_{0}^{2}}.
\end{align}
Consequently, this formulation does not exhibit tachyonic behavior; it describes an inhomogeneous Klein-Gordon equation with a source term. While this approach may be relevant for equilibrium configurations or gauge-fixed formulations, it does not provide a viable quantum mechanical framework for tachyonic Carrollian particles, as it lacks complex energy eigenvalues and exponential time evolution.

We therefore adopt Approach 1 as the physically meaningful and mathematically consistent framework for hybrid Carrollian quantum mechanics.

%%---------------------------------------%%
\subsection{Complex wavefunction and continuity equation}

For a complex Carrollian field, the hybrid equation takes the Hermitian form
\begin{align}
    \left(\hbar^{2}\partial_{t}^{2}
    - \hbar^{2}\partial_{x}^{2}
    - \hbar^{2}\chi'\partial_{t}
    + m_{0}^{2}\right)\psi = 0,
    \qquad \chi' \in \mathbb{R},
\end{align}
where \(\chi = i\chi'\) ensures \(U(1)\) invariance. The conjugate equation is
\begin{align}
    \left(\hbar^{2}\partial_{t}^{2}
    - \hbar^{2}\partial_{x}^{2}
    - \hbar^{2}\chi'\partial_{t}
    + m_{0}^{2}\right)\psi^{*} = 0.
\end{align}

Multiplying the first equation by \(\psi^{*}\), the second by \(\psi\), and subtracting yields the continuity relation
\begin{align}
    \hbar^{2}\partial_t(\psi^{*}\partial_t\psi - \psi\partial_t\psi^{*})
    - \hbar^{2}\partial_x(\psi^{*}\partial_x\psi - \psi\partial_x\psi^{*})
    - \hbar^{2}\chi'(\psi^{*}\partial_t\psi - \psi\partial_t\psi^{*}) = 0.
    \label{eq:hybrid_continuity_derived}
\end{align}

Solving this equation using the integrating factor method, the hybrid continuity equation takes the canonical form
\begin{align}
    \partial_t \rho + \nabla\cdot J = 0,
\end{align}
with conserved probability density and current
\begin{align}
    \rho 
    &= i\hbar^{2} e^{-\chi' t}
    \left(\psi^{*}\partial_t\psi - \psi\partial_t\psi^{*}\right), \label{eq:hybrid_rho_final} \\
    J 
    &= -i\hbar^{2} e^{-\chi' t}
    \left(\psi^{*}\partial_x\psi - \psi\partial_x\psi^{*}\right). \label{eq:hybrid_J_final}
\end{align}
For \(\chi' = 0\), these expressions reduce to the standard Klein-Gordon density and current.

The density \(\rho\) represents the probability density in the hybrid Carrollian sector. It generalizes the time-like sector probability density by incorporating the non-unitary contribution through the exponential factor \(e^{-\chi' t}\). This factor is essential for restoring conservation in the presence of the first-order temporal derivative, effectively absorbing the non-unitary dynamics into the definition of the probability density.

\subsection*{Overview}

The hybrid sector is mathematically well-defined and fully Carroll-invariant, though its physical interpretation is context-dependent. It may describe effective dynamics on non-stationary null surfaces or provide a quantum framework for fractonic systems that interpolate between immobile and mobile excitations. Future work should clarify whether the hybrid equation arises from a null reduction of a higher-dimensional Lorentzian theory or as an effective description near quantum critical points in condensed matter.

In the following subsections, we analyze the hybrid Carrollian particle in a box, study spatial and temporal tunneling, and discuss the unique dynamical features of this mixed Carrollian regime, distinguishing between real and complex wavefunction cases.

%----------------------------------------------------------------------
%----------------------------------------------------------------------
\subsection{Hybrid Carrollian particle in a box}\label{Hybrid Carrollian particle in a box}

Consider a hybrid Carrollian quantum particle confined to a \(1+1\)-dimensional infinite potential well:
\begin{align}
    V(x,t) = 
    \begin{cases}
        0, & 0 < x < L,\, 0 < t < T, \\
        \infty, & \text{otherwise}.
    \end{cases}
\end{align}

The hybrid Carrollian equation in the interior region is
\begin{align*}
    \big(\hbar^{2}\partial_{t}^{2}- \hbar^{2}\partial_{x}^{2}+ \hbar^{2}\chi\,\partial_{t} + m_{0}^{2}\big)\,\psi(x,t) = 0,
\end{align*}
where \(\chi \in \mathbb{R}\) is the damping parameter. Separating variables with the ansatz \(\psi(x,t)=\phi(x)\xi(t)\) yields  
\begin{align}
    \frac{\hbar^{2}\partial_{t}^{2}\,\xi(t)}{\xi(t)}+\frac{\hbar^{2}\chi\,\partial_{t}\,\xi(t)}{\xi(t)}+m_{0}^{2}
    \;=
    \frac{\hbar^{2}\partial_{x}^{2}\phi(x)}{\phi(x)}=-\lambda_{n}^{2},
\end{align}
where \(\lambda\) is the separation constant. This leads to the spatial eigenvalue problem  
\begin{align}
    \partial_{x}^{2}\phi(x)+\frac{\lambda_{n}^{2}}{\hbar^{2}}\phi(x)=0,
\end{align}
and the temporal equation  
\begin{align}
    \partial_{t}^{2}\xi(t)+\chi\,\partial_{t}\xi(t)+\frac{\lambda_{n}^{2}+m_{0}^{2}}{\hbar^{2}}\xi(t)=0.
\end{align}	

The solutions \(\phi_{n}(x)\) are the usual sine modes of the infinite well, while \(\xi_{n}(t)\) are damped oscillators whose frequencies are determined by the hybrid Carrollian parameters \(\chi\) and \(m_{0}\).

\paragraph{Temporal sector.}
The general temporal solution depends on the sign of \(\Omega_{n}^{2}\), defined as
\begin{align}
    \Omega_{n}^{2} = \frac{\chi^{2}}{4} - \omega_0^2, \qquad 
    \omega_0^{2} = \frac{\lambda_{n}^{2} + m_0^{2}}{\hbar^{2}}.
\end{align}
The sign of \(\Omega_{n}^{2}\) determines three distinct dynamical regimes:

\begin{enumerate}

    \item \textbf{$\Omega_{n}^{2} > 0$ (overdamped)}:  
    The two characteristic exponents are real and generically distinct. The temporal mode takes the form
    \begin{align}
        \xi(t) = e^{-\chi t/2}\left( A\, e^{\Omega_n t} + B\, e^{-\Omega_n t} \right).
    \end{align}
    No oscillations occur: for \(\chi > 0\), the amplitude decays monotonically; for \(\chi < 0\), it grows monotonically. This behavior is analogous to a heavily damped mechanical oscillator when \(\chi > 0\), and to an amplifying mode when \(\chi < 0\). 

    \item \textbf{$\Omega_{n}^{2}=0$ (critically damped)}:  
    At the critical point, the two characteristic roots coincide, yielding
    \begin{align}
        \xi(t) = (\alpha\, t + \beta)\, e^{-\chi t/2}.
    \end{align}
    The evolution is purely monotonic: for \(\chi > 0\), the amplitude decays at the fastest possible non-oscillatory rate; for \(\chi < 0\), it grows at the fastest non-oscillatory rate. This regime sits precisely between the underdamped and overdamped cases. The linear prefactor \(\alpha t\) reflects the fact that the system can no longer complete even a single oscillation before damping (or anti-damping) dominates. The characteristic relaxation (or divergence) time is \(\tau = 2/|\chi|\).

    \item \textbf{$\Omega_{n}^{2} < 0$ (underdamped)}:  
    When \(\Omega_{n}^{2} < 0\), the effective frequency becomes imaginary. Introducing the real oscillation frequency
    \begin{align}
        \omega_n = \sqrt{-\Omega_{n}^{2}}
        = \sqrt{\omega_0^{2} - \frac{\chi^{2}}{4}} > 0,
    \end{align}
    the temporal mode is
    \begin{align}
        \xi(t) = e^{-\chi t/2}\bigl( C \cos(\omega_n t) + D \sin(\omega_n t) \bigr).
    \end{align}
    This describes harmonic oscillations at frequency \(\omega_n\) with an exponential envelope. For \(\chi > 0\), the oscillations are damped; for \(\chi < 0\), the envelope grows, corresponding to amplification. This underdamped regime is characteristic of the hybrid Carrollian sector, where oscillatory and dissipative (or gain) behavior coexist, closely paralleling a damped harmonic oscillator.

\end{enumerate}

\paragraph{Spatial sector.}
The separation constant \(\lambda_n\) determines the spatial wave number. Defining
\begin{align}
    k_n^2 = \frac{\lambda_n^2}{\hbar^2},
\end{align}
the spatial part reduces to the Helmholtz equation
\begin{align}
    \phi''(x) + k_n^2 \phi(x) = 0.
\end{align}
Its general solution is a linear combination of sinusoidal modes:
\begin{align}
    \phi_n(x) = B_n \sin(k_n x) + C_n \cos(k_n x).
\end{align}
For a particle confined to an infinite potential well on the interval \(0 < x < L\), the wavefunction must vanish at the impenetrable boundaries:
\begin{align}
    \phi_n(0) = 0, \qquad \phi_n(L) = 0.
\end{align}
The condition \(\phi_n(0)=0\) forces \(C_n = 0\). Then \(\phi_n(L) = B_n \sin(k_n L) = 0\) requires \(\sin(k_n L)=0\), giving
\begin{align}
    k_n L = n\pi, \qquad n = 1,2,3,\dots
\end{align}
Thus the wave numbers are quantized:
\begin{align}
    k_n = \frac{n\pi}{L}.
\end{align}
The corresponding spatial eigenfunctions are the familiar sine modes:
\begin{align}
    \phi_n(x) = B_n \sin\!\left(\frac{n\pi x}{L}\right).
\end{align}
Substituting back into the definition of \(k_n\) gives the eigenvalues of the separation constant:
\begin{align}
    \lambda_n^2 = \hbar^2 k_n^2 = \hbar^2 \left(\frac{n\pi}{L}\right)^2.
\end{align}

The spatial part of the hybrid Carrollian particle in a box is identical to that of a non-relativistic quantum particle in an infinite well. The wave numbers are quantized because the wavefunction must fit into the box with nodes at the boundaries. This spatial quantization is independent of the damping parameter \(\chi\) and depends only on the box size \(L\). The eigenvalues \(\lambda_n^2\) determine the natural frequencies \(\omega_0\) that enter the temporal damped oscillator equation. Thus, the hybrid sector inherits the familiar spatial quantization of the infinite well while adding dissipative (or amplifying) time evolution through the parameter \(\chi\).
%-----------------------------------------------------------------------
%-----------------------------------------------------------------------
%%%%%%%%%%%%%%%%%%%%%%%%%%%%%%%%%%%%%%%%%%%%%%%%%%%%%%%%%%%%%%%%%%%%%%%%%%%%%%%%%%%%%%%%%%%%%%%%%%%%%%%%%%%%%%%%%%%%%%%%%%%%%%%%%%%%%%%%%%%%%%%%
\subsection{Hybrid Carrollian tunneling}

We now study tunneling phenomena in the hybrid Carrollian sector, where both temporal and spatial potentials may be present. The hybrid wave equation in $(1+1)$ dimensions is
\begin{align}
    \Bigl(
        \hbar^{2}\partial_{t}^{2}
        - \hbar^{2}\partial_{x}^{2}
        + \hbar^{2}\chi\,\partial_{t}
        + m_{0}^{2}
        + U(t)^{2} + U(x)^{2}
    \Bigr)\psi(x,t) = 0,
\end{align}
with $U(t)$ and $U(x)$ taken as rectangular barriers:
\begin{align}
    U(t) &=
    \begin{cases}
        0, & t < t_{1},\\
        V_{t}, & t_{1} \le t \le t_{2},\\
        0, & t > t_{2},
    \end{cases}
    \qquad
    U(x) =
    \begin{cases}
        0, & x < x_{1},\\
        V_{x}, & x_{1} \le x \le x_{2},\\
        0, & x > x_{2}.
    \end{cases}
\end{align}
These temporal and spatial potentials may act either separately or simultaneously on hybrid Carrollian particles. In the present analysis, we treat them independently, thereby covering three distinct physical situations:
\begin{itemize}
    \item \textbf{Purely spatial barrier} ($V_{t}=0$, $V_{x}\neq0$): the particle encounters a static spatial barrier, analogous to ordinary quantum tunneling but modified by the damping/amplification induced by the $\chi$ term.

    \item \textbf{Purely temporal barrier} ($V_{x}=0$, $V_{t}\neq0$): the potential changes discontinuously in time, mixing positive and negative frequency modes (temporal tunneling) as in the time-like Carroll sector; the spatial Laplacian and the $\chi$ term remain present and influence the evolution.

    \item \textbf{Combined barrier} ($V_{t}\neq0$, $V_{x}\neq0$): both effects coexist, producing a richer scattering problem in which the wavefunction may be partially reflected in space and partially converted in frequency.
\end{itemize}

Because the equation is linear and the potentials are piecewise constant, the general solution in each region can be constructed by separation of variables using the ansatz $\psi(x,t)=\phi(x)\,\xi(t)$. The temporal and spatial parts are decoupled by a separation constant (denoted $-\lambda^{2}$), and the matching conditions at the interfaces $x=x_{1},x_{2}$ and $t=t_{1},t_{2}$ determine the corresponding scattering amplitudes. In the following subsections we analyze the spatial and temporal barriers separately; the combined case can be obtained by successive application of the same matching procedure.

%----------------------------------------------------------------------
\subsubsection{Spatial potential}

When the potential depends only on the spatial coordinate, the spatial part of the problem splits into three distinct regions. The hybrid parameter \(\chi\) introduces damping (or amplification) in time, but because the temporal factor \(\xi(t)\) is common to all regions, it cancels in the matching conditions. Consequently, the spatial scattering probabilities are identical to those of the space-like Carroll sector. The temporal envelope \(e^{-\chi t/2}\) modifies the overall probability amplitude over time but does not affect the reflection and transmission coefficients. We consider a rectangular barrier of height \(V\) and width \(L = x_2 - x_1\):
\begin{align}
	U(x) = 
	\begin{cases}
		0, & x < x_1,\\
		V, & x_1 \le x \le x_2,\\
		0, & x > x_2.
	\end{cases}
\end{align}
Separating variables \(\psi(x,t)=\phi(x)\xi(t)\) and using the separation constant \(-\lambda^2\) (with \(\lambda^2>0\)) yields the spatial equation
\begin{align}
	-\hbar^2 \phi''(x) + U(x)^2 \phi(x) = -\lambda^2 \phi(x),
\end{align}
which can be rewritten as
\begin{align}
	\phi''(x) = \frac{U(x)^2 - \lambda^2}{\hbar^2}\,\phi(x).
\end{align}
In the free regions (\(U=0\)) this becomes \(\phi'' = -(\lambda^2/\hbar^2)\phi\). Hence we define the wave number
\begin{align*}
	k = \frac{\lambda}{\hbar},
\end{align*}
so that \(\phi'' + k^2 \phi = 0\). The general solution is a superposition of plane waves.

Inside the barrier (\(U=V\)), we have \(\phi'' = (V^2 - \lambda^2)/\hbar^2 \,\phi\). For a barrier we require \(\lambda^2 < V^2\) so that the coefficient is positive. Define
\begin{align*}
	\kappa = \frac{\sqrt{V^2 - \lambda^2}}{\hbar} > 0,
\end{align*}
then \(\phi'' = \kappa^2 \phi\), and the solutions are exponential: \(e^{\pm \kappa x}\). The regional solutions are therefore:
\begin{itemize}
	\item \textbf{Region $I$} (\(x < x_1\)): \(\phi_I(x) = A e^{ikx} + B e^{-ikx}\).
	\item \textbf{Region $II$} (\(x_1 \le x \le x_2\)): \(\phi_{II}(x) = C e^{\kappa x} + D e^{-\kappa x}\).
	\item \textbf{Region $III$} (\(x > x_2\)): \(\phi_{III}(x) = F e^{ikx}\) (no incoming wave from the right, \(G=0\)).
\end{itemize}

\subsubsection*{Matching conditions and solution}

Continuity of \(\phi\) and \(\phi'\) at \(x=x_1\) and \(x=x_2\) yields four equations. Shifting the origin to \(x_1\) (\(y=x-x_1\)) so that the barrier occupies \(0 \le y \le L\), the linear system is solved to give the transmission and reflection amplitudes:
\begin{align}
	\mathcal{T} &\equiv \frac{F}{A} = \frac{4k\kappa\,e^{-ikL}}{(k+\kappa)^2 e^{\kappa L} - (k-\kappa)^2 e^{-\kappa L}},\\
	\mathcal{R} &\equiv \frac{B}{A} = \frac{(k^2-\kappa^2)\sinh(\kappa L)}{2ik\kappa\cosh(\kappa L) + (k^2+\kappa^2)\sinh(\kappa L)}.
\end{align}
The corresponding probabilities are \(T = |\mathcal{T}|^2\) and \(R = |\mathcal{R}|^2 = 1-T\). 

The hybrid Carrollian particle tunnels through a spatial barrier with exactly the same probability as a non-relativistic Schrödinger particle. The hybrid nature manifests only in the time domain: the temporal part contributes a factor \(e^{-\chi t/2}\) (common to all regimes, underdamped or overdamped), so the probability amplitude decays uniformly (if \(\chi>0\)) or grows (if \(\chi<0\)) in time. Thus, the reflection and transmission ratios are constant in time, but the overall intensity changes. This behavior is absent in the pure space-like sector, where the time evolution is constant (or exponential only for the separation constant \(\lambda=0\)). Note that the mass \(m_0\) does not appear in the spatial equation; it only enters the temporal damped oscillator through the frequency \(\omega_0^2 = (m_0^2+\lambda^2)/\hbar^2\).
%----------------------------------------------------------------------%----------------------------------------------------------------------%----------------------------------------------------------------------
\subsubsection{Temporal potential}

The analysis now turns to the temporal potential. In this context, the sign of \(\Omega^{2}\) plays a central role, as already noted in Sec.~\ref{Hybrid Carrollian particle in a box}. 

When analyzing a purely temporal potential barrier \(U(t)\) in the hybrid Carrollian sector, the separation constant \(\lambda^2 = \hbar^2 k^2\) (where \(k = |\vec{k}|\) is the spatial wave number) appears in the temporal equation. The wavefunction has a non-trivial spatial profile, e.g., a plane wave \(\phi(\vec{x}) = e^{i\vec{k}\cdot\vec{x}}\) with definite momentum \(\hbar\vec{k}\). The temporal equation becomes
\begin{align}
    \ddot{\xi} + \chi \dot{\xi} + \frac{m_0^2 + \hbar^2 k^2 + U(t)^2}{\hbar^2}\,\xi = 0.
\end{align}
The spatial momentum contributes an effective mass term \(\hbar^2 k^2\) that adds to \(m_0^2\). Consequently, different Fourier modes experience different effective frequencies inside the barrier, and the temporal tunneling amplitudes acquire a \(k\) dependence.

For clarity, we define the quantities that control the damping regimes. In the free regions (\(U(t)=0\)),
\[
\Omega_0^2 = \frac{\chi^2}{4} - \frac{m_0^2 + \hbar^2 k^2}{\hbar^2},
\]
and inside the temporal barrier (\(U(t)=V_t\)),
\[
\Omega^2 = \frac{\chi^2}{4} - \frac{m_0^2 + \hbar^2 k^2 + V_t^2}{\hbar^2}.
\]

The sign of \(\Omega_0^2\) (respectively \(\Omega^2\)) determines whether the free (barrier) region is overdamped (\(\Omega^2 > 0\)), critically damped (\(\Omega^2 = 0\)), or underdamped (\(\Omega^2 < 0\)). Depending on the signs in the free regions and inside the barrier, three distinct physical regimes can be identified:

\paragraph{Case 1: Overdamped free regions, underdamped barrier.}
In this scenario the free regions (\(I\) and \(III\)) are overdamped with \(\Omega_0^2 > 0\), while the barrier region (\(II\)) is underdamped with \(\Omega^2 < 0\). The solutions are
\begin{align}
    \xi_I(t) &= e^{-\chi t/2}\bigl(A e^{\Omega_0 t} + B e^{-\Omega_0 t}\bigr),\\
    \xi_{II}(t) &= e^{-\chi t/2}\bigl(C \cos(\omega t) + D \sin(\omega t)\bigr), \quad \omega = \sqrt{-\Omega^2},\\
    \xi_{III}(t) &= e^{-\chi t/2}\bigl(E e^{\Omega_0 t} + F e^{-\Omega_0 t}\bigr).
\end{align}

\paragraph{Case 2: Critically damped free regions, underdamped barrier.}
Here the free regions are critically damped (\(\Omega_0 = 0\)) and the barrier is underdamped (\(\Omega^2 < 0\)). The solutions read
\begin{align}
    \xi_I(t) &= e^{-\chi t/2}(A + B t),\\
    \xi_{II}(t) &= e^{-\chi t/2}\bigl(C \cos(\omega t) + D \sin(\omega t)\bigr),\\
    \xi_{III}(t) &= e^{-\chi t/2}(E + F t).
\end{align}

\paragraph{Case 3: Underdamped free regions, underdamped barrier.}
In this case all three regions are underdamped (\(\Omega_0^2 < 0\) and \(\Omega^2 < 0\)). The solutions are oscillatory everywhere:
\begin{align}
    \xi_I(t) &= e^{-\chi t/2}\bigl(A \cos(\omega_0 t) + B \sin(\omega_0 t)\bigr), \quad \omega_0 = \sqrt{-\Omega_0^2},\\
    \xi_{II}(t) &= e^{-\chi t/2}\bigl(C \cos(\omega t) + D \sin(\omega t)\bigr), \quad \omega = \sqrt{-\Omega^2},\\
    \xi_{III}(t) &= e^{-\chi t/2}\bigl(E \cos(\omega_0 t) + F \sin(\omega_0 t)\bigr).
\end{align}

For any of these regimes, the matching conditions at \(t=t_1\) and \(t=t_2\) (continuity of \(\xi\) and \(\dot{\xi}\)) determine the scattering amplitudes. The structure is analogous to the time-like Carroll case discussed in Sec.~\ref{subsec:timelike_tunneling}. In the underdamped case, for instance, the matching yields a Bogoliubov transformation between the incoming and outgoing modes, with the well-known relation \(|\mathcal{T}|^2 - |\mathcal{R}|^2 = 1\), reflecting the indefinite norm associated with the second-order time derivatives. The explicit amplitudes can be obtained by following the same steps as in Sec.~\ref{subsec:timelike_tunneling}, with the replacement \(m_0 \to \sqrt{m_0^2 + \hbar^2 k^2}\) and incorporating the common exponential factor \(e^{-\chi t/2}\), which cancels in the matching. We omit the full algebraic expressions here, as they are straightforward generalizations of the time-like calculation.

\paragraph{Concrete implementation for the underdamped barrier (Case 1).}
We now specialize to the physically most relevant situation: the free regions are overdamped (\(\Omega_0^2 > 0\)) and the barrier is underdamped (\(\Omega^2 < 0\)). For clarity, we set \(k = 0\) in this implementation.

Define the real oscillation frequency inside the barrier:
\[
\omega = \sqrt{-\Omega^2} = \sqrt{\frac{m_0^2 + V^2}{\hbar^2} - \frac{\chi^2}{4}} \; (>0).
\]
For the free regions we keep \(\Omega_0 = \sqrt{\chi^2/4 - m_0^2/\hbar^2} > 0\).  
The temporal solutions are then
\begin{align}
    \xi_I(t) &= e^{-\chi t/2}\bigl(A e^{\Omega_0 t} + B e^{-\Omega_0 t}\bigr), \quad t < t_1,\\
    \xi_{II}(t) &= e^{-\chi t/2}\bigl(C \cos(\omega t) + D \sin(\omega t)\bigr), \quad t_1 \le t \le t_2,\\
    \xi_{III}(t) &= e^{-\chi t/2}\bigl(E e^{\Omega_0 t} + F e^{-\Omega_0 t}\bigr), \quad t > t_2.
\end{align}
We consider an incident wave coming from the left (earlier times). For the solution to remain finite as \(t \to -\infty\), we must set the growing exponential coefficient to zero, hence \(B = 0\). The coefficient \(A\) is then the incident amplitude.

\paragraph{Matching at \(t = t_1\).}
Continuity of \(\xi\) and \(\dot{\xi}\) at the left interface gives
\begin{align}
    A e^{\Omega_0 t_1} &= C\cos(\omega t_1) + D\sin(\omega t_1), \label{eq:match1}\\
    \Omega_0 A e^{\Omega_0 t_1} &= -\frac{\chi}{2}\bigl(C\cos(\omega t_1) + D\sin(\omega t_1)\bigr) + \omega\bigl(-C\sin(\omega t_1) + D\cos(\omega t_1)\bigr). \label{eq:match2}
\end{align}
Substituting \eqref{eq:match1} into \eqref{eq:match2} and solving for \(C\) and \(D\) yields
\begin{align}
    C &= A e^{\Omega_0 t_1}\Bigl[\cos(\omega t_1) - \frac{\Omega_0}{\omega}\sin(\omega t_1)\Bigr],\\
    D &= A e^{\Omega_0 t_1}\Bigl[\sin(\omega t_1) + \frac{\Omega_0}{\omega}\cos(\omega t_1)\Bigr].
\end{align}

\paragraph{Matching at \(t = t_2\).}
At the right interface,
\begin{align}
    C\cos(\omega t_2) + D\sin(\omega t_2) &= E e^{\Omega_0 t_2} + F e^{-\Omega_0 t_2}, \label{eq:match3}\\
    -\omega C\sin(\omega t_2) + \omega D\cos(\omega t_2) &= \Omega_0\bigl(E e^{\Omega_0 t_2} - F e^{-\Omega_0 t_2}\bigr). \label{eq:match4}
\end{align}
Solving the linear system \eqref{eq:match3}-\eqref{eq:match4} for \(E\) and \(F\) gives the outgoing amplitudes:
\begin{align}
    E &= \frac{e^{-\Omega_0 t_2}}{2}\Bigl[ C\cos(\omega t_2) + D\sin(\omega t_2) 
    + \frac{\omega}{\Omega_0}\bigl(-C\sin(\omega t_2) + D\cos(\omega t_2)\bigr) \Bigr],\\[6pt]
    F &= \frac{e^{\Omega_0 t_2}}{2}\Bigl[ C\cos(\omega t_2) + D\sin(\omega t_2) 
    - \frac{\omega}{\Omega_0}\bigl(-C\sin(\omega t_2) + D\cos(\omega t_2)\bigr) \Bigr].
\end{align}
These expressions determine the transmitted (\(E\)) and reflected (\(F\)) amplitudes in terms of the incident amplitude \(A\) (since \(C\) and \(D\) are already expressed through \(A\)).

%%%%%%%%%%%%%%%%%%%%%%%%%%%%%%%%%%%%%%%%%%%%%%%%%%%%%%%%%%%%%%%%%%%%%%%%%%%%%%%%%%%%%%%%%%%%%%%%%%%%%%%%%%%%%%%%%%%%%%%%%%%%%%%%%%%%%%%%%%%%%%%%
%%%%%%%%%%%%%%%%%%%%%%%%%%%%%%%%%%%%%%%%%%%%%
\section{Conclusion}\label{conclusion}

In this work we present the first unified and systematic formulation of Carrollian quantum mechanics, derived from controlled ultra-relativistic limits of the Klein-Gordon equation. By analyzing the $c\to0$ limit in full generality, we demonstrate that the Carrollian regime does not yield a single quantum theory but instead splits into three inequivalent sectors---time-like, space-like, and hybrid---each with its own wave equation, continuity equation, probability current, and physical interpretation.

\paragraph{Summary of achievements.}
\begin{itemize}
    \item \textbf{Time-like Carroll QM}: Only temporal derivatives survive,  leading to ultra-local dynamics with vanishing spatial current. We discovered  a genuinely new phenomenon \emph{temporal tunneling} in which a time-dependent potential mixes positive and negative frequencies, resulting in a Bogoliubov transformation with $|\mathcal{T}|^2=1+|\mathcal{R}|^2$. We solved the  temporal box problem, obtaining a discrete frequency spectrum and mass  quantization, and clarified that the uncertainty principle remains valid  despite the absence of spatial propagation.
    
   \item \textbf{Space-like Carroll QM}: Spatial propagation is retained while temporal evolution becomes first-order. The sector admits two distinct Carroll-invariant formulations. In the coupled formulation (Approach 1), a compensating field $\chi$ restores Carroll boost invariance and yields purely imaginary energy eigenvalues, signaling tachyonic behavior with null states. In the additive formulation (Approach 2), the equation reduces to a static constraint devoid of dynamics, describing non-tachyonic configurations. We constructed the probability current, solved the infinite well, and analyzed conventional spatial tunneling through a rectangular barrier. The physical content of the tachyonic sector resides in the spatial current rather than in the density, reflecting the null nature of the states.
    
    \item \textbf{Hybrid Carroll QM}: This mixed sector retains both temporal and spatial derivatives and admits two distinct Carroll-invariant formulations. In the coupled formulation (Approach 1), the equation yields a damped or amplified wave equation with complex energy eigenvalues for complex wavefunctions, signaling tachyonic behavior, and reduces to a damped harmonic oscillator with real coefficients for real wavefunctions. In the additive formulation (Approach 2), the equation reduces to a non-tachyonic inhomogeneous Klein-Gordon equation with real energy spectrum. We classified the particle in a box solutions into underdamped, critically damped, and overdamped regimes, and solved both spatial and temporal barrier scattering. The probability density takes a modified form, incorporating an exponential factor that ensures conservation through the continuity equation, providing a consistent probabilistic interpretation.
\end{itemize}
For each sector we provided a consistent probabilistic interpretation (continuity equation, probability density, and current), exact solutions for confinement and scattering, and clear physical explanations of the novel Carrollian effects.

\paragraph{Future directions.}
The framework developed here opens several avenues for further research, of which we highlight three particularly promising directions:
\begin{itemize}
    
    \item \textbf{Quantum information and complexity}: The temporal barrier acts as a tunable beam splitter for frequency modes, capable of generating entangled pairs of modes. This suggests applications in quantum information, including immobile qubits that interact via time-dependent couplings and the study of quantum complexity in non-Lorentzian settings.
    
    \item \textbf{Hybrid Carrollian quantum mechanics and Carroll swiftons}:
    A natural direction for future work is a deeper investigation of the relationship between hybrid Carrollian quantum mechanics and Carroll swiftons. The structural parallels between these systems---most notably their shared mixed temporal---spatial dynamics and the presence of compensating mechanisms linking temporal evolution to spatial variation--- suggest that hybrid CQM may provide the quantum-mechanical underpinning of  swifton-type excitations. Conversely, swifton field theories offer a natural higher-dimensional completion of the hybrid sector, including geometric back-reaction through Carrollian torsion. Clarifying this correspondence---at the level of dispersion relations and through explicit  couplings to Carroll gravity---may reveal a unified framework connecting quantum Carroll systems to their field-theoretic and geometric counterparts.
    
    \item \textbf{Galilean quantum mechanics from similar contractions}: The method used here---controlled In\"on\"u--Wigner contractions of the Klein-Gordon equation---can be adapted to the Galilean (non-relativistic)  limit $(c\to\infty)$. A systematic derivation of the three Galilean sectors (time-like, space-like, and hybrid) would provide a complete quantum  mechanical counterpart to the well-known Galilean symmetry, complementing the Schr\"odinger equation and its extensions. This would unify the treatment of both ultra-relativistic (Carroll) and non-relativistic (Galilei)  limits within the same framework. Following the methods of this work, one could solve the particle in a box and rectangular barrier problems for all  Galilean sectors, uncover novel effects analogous to temporal tunneling  (e.g., "spatial tunneling with Galilean dispersion"), and establish dualities with fracton phases in the non-relativistic regime, thereby  filling a gap parallel to the Carrollian case.
\end{itemize}

%%%%%%%%%%%%%%%%%%%%%%%%%%%%%%%%%%%%%%%%%%%%%%%%%%%%%%%%%%%%%%%%%%%%%%%%%%%%%%%%%%%%%%%%%%%%%%%%%%%%%%%%%%%%%%%%%%%%%%%%%%%%%%%%%%%%%%%%%%%%%%%%%%%%%%%%%%%%%%%%%%%%%%%%%%%%%%%%%%%%%%%%%%%%%%%%%%%%%%%%%%%%%%%%%%%%%%%%%%%%%%%%%%%%%%%%%%%%%%%%%%%%%%%%%%%%%%%%%%%%%%%%%%%%%%%%%%%%%%%%%%%%%%%%%%%%%%%%%%%%%%%%%%%%%%%%%%%%%%%%%%%%%%%%%%%%%%%%%%%%%%%%%%%%%%%%%%%%%%%%%%%%%%
\acknowledgments

I am grateful to Hamid Afshar, Shahin Sheikh-Jabbari, Bahram Shakerin, Ali Parvizi, Vahid Taghiloo, Ida Rasolian, Mojtaba Najafizadeh, and Alireza Pahlavan for their insightful contributions and early discussions on Carrollian quantum mechanics.

%%%%%%%%%%%%%%%%%%%%%%%%%%%%%%%%%%%%%%%%%%%%%%%%%%%%%%%%%%%%%%%%%%%%%%%%%%%%%%%%%%%%%%%%%%%%%%%%%%%%%%%%%%%%%%%%%%%%%%%%%%%%%%%%%%%%%%%%%%%%%%%%%%%%%%%%%%%%%%%%%%%%%%%%%%%%%%%%%%%%%%%%%%%%%%%%%%%%%%%%%%%%%%%%%%%%%%%%%%%%%%%%%%%%%%%%%%%%%%%%%%%%%%%%%%%%%%%%%%%%%%%%%%%%%%%%%%%%%%%%%%%%%%%%%%%%%%%%%%%%%%%%%%%%%%%%%%%%%%%%%%%%%%%%%%%%%%%%%%%%%%%%%%%%%%%%%%%%%%%%%%%%%%
\appendix

%%%%%%%%%%%%%%%%%%%%%%%%%%%%%%%%%%%%%%%%%%%%%%%%%%%%%%%%%%%%%%%%%%%%%%%%%%%%%%%%%%%%%%%%%%%%%%%%%%%%%%%%%%%%%%%%%%%%%%%%%%%%%%%%%%%%%%%%%%%%%%%%%%%%%%%%%%%%%%%%%%%%%%%%%%%%%%%%%%%%%%%%%%%%%%%%%%%%%%%%%%%%%%%%%%%%%%%%%%%%%%%%%%%%%%%%%%%%%%%%%%%%%%%%%%%%%%%%%%%%%%%%%%%%%%%%%%%%%%%%%%%%%%%%%%%%%%%%%%%%%%%%%%%%%%%%%%%%%%%%%%%%%%%%%%%%%%%%%%%%%%%%%%%%%%%%
%%%%%%%%%%%%%%%%%%%%%%%%%%%%%%%%%%%%%%%%%%%%%%%%%%%%%%%%%%%%%%%%%%%%%%%%%%%%%%%%%%%%%%%%%%%%%%%%%%%%%%%%%%%%%%%%%%%%%%%%%%%%%%%%%%%%%%%%%%%%%%%%%%%%%%%%%%%%%%%%%%%%%%%%%%%%%%%%%%%%%%%%%%%%%%%%%%%%%%%%%%%%%%%%%%%%%%%%%%%%%%%%%%%%%%%%%%%%%%%%%%%%%%%%%%%%%%%%%%%%%%%%%%%%%%%%%%%%%%%%%%%%%%%%%%%%%%%%%%%%%%%%%%%%%%%%%%%%%%%%%%%%%%%%%%%%%%%%%%%%%%%%%%%%%%%%%%%%%%%%%%%%%%
\section{Classical Hybrid limit and particles}\label{app:classical_hybrid}

In the main text we derived the quantum mechanical hybrid sector from a controlled 
contraction of the Klein-Gordon equation. Here we show that a classical analogue 
of the hybrid Carroll particle exists and can be obtained by a consistent scaling 
of the relativistic point particle action. This provides a classical counterpart 
to the quantum hybrid sector and clarifies the role of the compensating field 
$\chi$.

\paragraph{Motivation.}
The relativistic point particle action in Hamiltonian form is 
\cite{Ahmadi-Jahmani:2025iqc}
\begin{align}\label{RPP}
    S = \int d\tau \left[ -p_\mu \dot{x}^\mu - \frac{e}{2}\bigl(p^2 - m^2\bigr) \right],
\end{align}
where $x^\mu(\tau)$ and $p^\mu(\tau)$ are the spacetime coordinates and conjugate 
momenta, $e(\tau)$ is the einbein (a Lagrange multiplier enforcing the mass-shell 
constraint), and the dot denotes differentiation with respect to the world-line 
parameter $\tau$. This action is invariant under reparameterization and under 
Poincaré transformations. To obtain a Carrollian limit we send the speed of 
light to zero, which in phase space amounts to a specific rescaling of 
coordinates and momenta.

\paragraph{Classical hybrid limit.}
We consider the following scaling (the "hybrid" contraction):
\begin{align}\label{classical_hybrid_limit}
    x^{0} \to \epsilon\, x^{0},\qquad
    p^{0} \to \frac{E}{\epsilon},\qquad
    \vec{x} \to \epsilon\,\vec{x},\qquad
    \vec{p} \to \frac{\vec{p}}{\epsilon},\qquad
    e \to \epsilon^{2}\, e,
\end{align}
where $\epsilon$ is a dimensionless parameter. In the ultra-relativistic limit 
$\epsilon\to 0$, the rescaled variables remain finite. Substituting these 
rescalings into the action \eqref{RPP} and discarding total derivatives, we 
obtain
\begin{align}\label{classical_hybrid_particles}
    S = \int d\tau \left[ -E\,\dot{t} + \vec{p}\cdot\dot{\vec{x}} - \frac{e}{2}\bigl(E^{2} - \vec{p}^{\,2} - m^{2}\bigr) \right].
\end{align}
Here we have identified $x^{0}=t$ and $\dot{t}=dx^{0}/d\tau$. This action is 
the starting point for a classical description of a hybrid Carroll particle.

\paragraph{Lack of Carroll invariance and the compensating field $\chi$.}
The action \eqref{classical_hybrid_particles} is not invariant under Carroll 
boosts, which act as
\[
t' = t - \vec{\beta}\cdot\vec{x},\qquad \vec{x}' = \vec{x},\qquad
E' = E,\qquad \vec{p}' = \vec{p} + \vec{\beta}E,
\]
for a boost parameter $\vec{\beta}$. Under these transformations the kinetic 
term $-E\dot{t} + \vec{p}\cdot\dot{\vec{x}}$ changes by a total derivative, 
but the constraint term $\frac{e}{2}(E^{2}-\vec{p}^{2}-m^{2})$ is not invariant 
because $\vec{p}^{2}$ transforms inhomogeneously. To restore invariance, we 
introduce a compensating field $\chi(\tau)$ that couples linearly to $E$:
\begin{align}
    S_{\text{inv}} = \int d\tau \left[ -E\,\dot{t} + \vec{p}\cdot\dot{\vec{x}} 
    - \frac{e}{2}\bigl(E^{2}-\vec{p}^{\,2}-m^{2}\bigr) - \chi E \right].
\end{align}
Under a Carroll boost, $\chi$ must transform as
\begin{align}
    \delta_C \chi = -e\,\vec{p}\cdot\vec{\beta},
\end{align}
which cancels the non-invariant terms coming from $\vec{p}^{\,2}$. The presence 
of $\chi$ thus ensures full Carroll invariance. The field $\chi$ is 
non-dynamical; it plays the role of a St\"uckelberg field and does not 
introduce additional physical degrees of freedom.

%%%%%%%%%%%%%%%%%%%%%%%%%%%%%%%%%%%%%%%%%%%%%%%%%%%%%%%%%%%%%%%%%%%%%%%%%%%%%%%%%%%%%%%%%%%%%%%%%%%%%%%%%%%%%%%%%%%%%%%%%%%%%%%%%%%%%%%%%%%%%%%%%%%%%%%%%%%%%%%%%%%%%%%%%%%%%%%%%%%%%%%%%%%%%%%%%%%%%%%%%%%%%%%%%%%%%%%%%%%%%%%%%%%%%%%%%%%%%%%%%%%%%%%%%%%%%%%%%%%%%%%%%%%%%%%%%%%%%%%%%%%%%%%%%%%%%%%%%%%%%%%%%%%%%%%%%%%%%%%%%%%%%%%%%%%%%%%%%%%%%%%%%%%%%%%%%%%%%%%%%%%%%%
\addcontentsline{toc}{section}{References}
\bibliographystyle{fullsort.bst}
\bibliography{biblio}

\end{document}